\documentclass[12pt]{article}
\usepackage{amsmath,amssymb,cite}
\usepackage{graphicx}
\usepackage[a4paper,twoside]{geometry}
\begin{document}
\author{{\bf Diana Duplij}\renewcommand{\thefootnote}{\arabic{footnote})} \footnotemark\\
  \emph{Institute of Molecular Biology and Genetics, 150 Zabolotny Str.}\\
\emph{Kiev 03143, Ukraine} \and {\bf Steven  Duplij}\renewcommand{\thefootnote}{\arabic{footnote})}
\footnotemark\\
\emph{Kharkov National University, 4 Svoboda. Sq.} \\\emph{Kharkov 61077, Ukraine}}
\title{{\bf DETERMINATIVE DEGREE AND NUCLEOTIDE SEQUENCE ANALYSIS BY TRIANDERS}}
\maketitle

\stepcounter{footnote}
\footnotetext{E-mail: duplijd@mail.ru}
\stepcounter{footnote}
\footnotetext{E-mail: Steven.A.Duplij@univer.kharkov.ua; WWW:
http://www.math.uni-mannheim.de/\~{}duplij}

\begin{abstract}
A new version of DNA walks, where nucleotides are regarded unequal in their contribution to a
walk is introduced, which allows us to study thoroughly the "fine structure" of nucleotide
sequences. The approach is based on the assumption that nucleotides have an inner abstract
characteristics, the determinative degree, which reflects phenomenological properties of genetic
code and is adjusted to nucleotides physical properties. We consider each position in codon
independently, which gives three separate walks being characterized by different angles and
lengths, and such an object is called triander which reflects the ``strength'' of branch. A
general method of identification of DNA sequence "by triander", which can be treated as a unique
"genogram", "gene passport" is proposed. The two- and three-dimensional trianders are
considered.\ The difference of sequences fine structure in genes and the intergenic space is
shown. A clear triplet signal in coding locuses is found which is absent in the intergenic space
and is independent from the sequence length. The topological classification of trianders is
presented which can allow us to provide a detail working out signatures of functionally
different genomic regions.
\end{abstract}

\section{Introduction}

The genomic DNA sequence analysis using wide range of statistical methods
\cite{tor/whi/xie,bul2,luo/lee/jia,str1,fic/tor/wol,buk/doc/gol/hav,azb1} and
various symmetry investigations
\cite{fin/fin/mcg,hor/hor,bas/tso/jar,bhr,for/sac1,fra/sci/sor} is an
extremely important tool in extracting hidden information about the dynamic
process of evolution, especially after the availability of fully sequenced
genomes \cite{nak/goj/ike}. One of the most promising approaches is the DNA
walks method \cite{ham3,gat2,ber/gla/sko} (firstly introduced by Azbel
\cite{azb}) or genomic landscapes \cite{lobr1}, which is based on mapping of a
sequence into one-, two- or multidimensional metric space according to
various specific rules. In brief, while drawing a DNA\ walk, the corresponding
mappings assign a direction/unit vector to each nucleotide, to dinucleotide or
to purine (pyrimidine). The resulting broken lines endow a visual presentation
to a formal sequence of 4 symbols, where inhomogeneous regions, fluctuations,
\textquotedblleft patches\textquotedblright\ etc. \cite{ber/olo/fil} are
imme\-diately seen. A modification of the DNA walks method deals with each
position in codons independently, which gives three separate broken lines
being characteri\-zed by different angles and lengths \cite{ceb/dud}, where
also addition and subtraction of DNA\ walks were considered \cite{kow/mac/mac}.

Here we introduce a new version of DNA\ walks, where all 4 nucleotides are
regarded unequal in the sense that they give contribution to a walk differing
not only by direction, but also by module. It follows from the assumption
\cite{e-dup/dup1} that nucleotides have an inner abstract characteristics ---
the determinative degree \cite{e-dup/dup0} which reflects phenomenological
properties of genetic code and is adjusted to nucleotides physical properties.

\section{Genetic code redundancy, doublet matrix inner structure and
determinative degree}

As is well-known, the genetic code is a highly organized system \cite{e-ycas}
and has several general properties: triplet character, uniqueness,
nonoverlapping, commaless, redundancy (degeneracy), which means that most
amino acids can be specified by more than one codon \cite{e-lewi,e-sten}.

From 64 possible codons one can extract 16 families each defined by first two
nucleotides. Let we denote a triplet (5'--1--2--3--3') by \textbf{XYZ}$.$ Then
the codon sense can be fully determined by first two nucleotides \textbf{X}
and \textbf{Y} independently of third $\mathbf{Z}$. There are 8 unmixed
families (all 4 codons encode the same amino acid), and 8 mixed families for
which several patterns of assignment exist, in 6 of the latter the pyrimidine
codons ($\mathbf{Z}=\mathbf{C,U}$) determine one amino acid, and the purine
codons ($\mathbf{Z}=\mathbf{A},\mathbf{G}$) determine other ones or
termination signals (in one family). It was found that two third part of all
DNA bases are identical for all organisms for the sake of first two
nucleotides in a triplet, and variability of DNA\ composition is given by the
third base \cite{e-sing/berg1,e-lewi}.

All 16 doublets $\mathbf{XY}$ can be presented as the canonical matrix
\cite{e-rum1}
\begin{equation}%
\begin{tabular}
[c]{ccccccc}\cline{4-4}
&  &  & \multicolumn{1}{|c}{$\mathbf{CC}$} & \multicolumn{1}{|c}{} &  &
\\\cline{3-3}\cline{5-5}
&  & \multicolumn{1}{|c}{$\mathbf{GC}$} &  & $\mathbf{CG}$ &
\multicolumn{1}{|c}{} & \\\cline{2-2}\cline{6-6}
& \multicolumn{1}{|c}{$\mathbf{CU}$} &  & $\mathbf{GG}$ &  & $\mathbf{CU}$ &
\multicolumn{1}{|c}{}\\\cline{1-1}\cline{3-4}\cline{6-7}%
\multicolumn{1}{|c}{$\mathbf{AC}$} &  & \multicolumn{1}{|c}{$\mathbf{UG}$} &
& \multicolumn{1}{|c}{$\mathbf{GU}$} & \multicolumn{1}{|c}{} &
\multicolumn{1}{c|}{$\mathbf{CA}$}\\\cline{1-2}\cline{5-5}\cline{7-7}
& \multicolumn{1}{|c}{$\mathbf{AG}$} &  & $\mathbf{UU}$ &  & $\mathbf{GA}$ &
\multicolumn{1}{|c}{}\\\cline{2-2}\cline{6-6}
&  & \multicolumn{1}{|c}{$\mathbf{AU}$} &  & $\mathbf{UA}$ &
\multicolumn{1}{|c}{} & \\\cline{3-3}\cline{5-5}
&  &  & \multicolumn{1}{|c}{$\mathbf{AA}$} & \multicolumn{1}{|c}{} &  &
\\\cline{4-4}%
\end{tabular}
\ \ \ \ \label{0}%
\end{equation}
called the \textquotedblleft rhombic code\textquotedblright%
\ \cite{e-kar/sor,e-kara1}. They are grouped together in 2 octets
distinguished by ability of amino acid determination: 8 doublets \textbf{CC},
\textbf{AC}, \textbf{GC}, \textbf{CU}, \textbf{GU}, \textbf{UC}, \textbf{CG},
\textbf{GG} determine amino acid independently of third base (upper part in
(\ref{0})), and so they can be called \textquotedblleft
strong\textquotedblright, and other 8 doublets \textbf{AA},\textbf{AU},
\textbf{UU}, \textbf{CA}, \textbf{GA}, \textbf{UG}, \textbf{AG, UA }(lower
part in (\ref{0})) for which third base determines content of codons can be
called \textquotedblleft weak\textquotedblright\ ones \cite{e-rum1,e-ratn}.
The \textquotedblleft strong\textquotedblright\ set of doublets has the
following relative content $\mathbf{C:G:U:A}=7:5:3:1$, while the
\textquotedblleft weak\textquotedblright\ set has the reverse content
$\mathbf{C:G:U:A}=1:3:5:7$ \cite{e-ratn2}. Note that there is only one
$\mathbf{A}$ in the \textquotedblleft strong\textquotedblright\ octet, and one
$\mathbf{C}$ in \textquotedblleft weak\textquotedblright\ octet, and all 4
doublets with $\mathbf{Y}=\mathbf{C}$ completely determine amino acid, but
only 2 doublets with $\mathbf{Y}=\mathbf{G}$ and $\mathbf{Y}=\mathbf{U}$
completely determine it, while doublets with $\mathbf{Y}=\mathbf{A}$ never
determine amino acid. Thus, 4 nucleotides can be arranged in descending order
$\mathbf{C}$\textbf{, }$\mathbf{G}$\textbf{, }$\mathbf{U}$\textbf{,
}$\mathbf{A}$ by their determinative ability (\textquotedblleft
strength\textquotedblright) \cite{e-rum1,e-rum3}.

We introduce a numerical characteristics of the empirical \textquotedblleft
strength\textquotedblright\ ---\ \textit{determinative degree} of nucleotide
$\mathbf{d}_{\mathbf{X}}$ in the following way%

\begin{equation}%
\begin{array}
[c]{cccc}%
\text{Pyrimidine} & \text{Purine} & \text{Pyrimidine} & \text{Purine}\\
\mathbf{C} & \mathbf{G} & \mathbf{T/U} & \mathbf{A}\\%
\begin{array}
[c]{c}%
\mathbf{d}_{\mathbf{C}}=\mathbf{4}\\%
\begin{array}
[c]{c}%
\text{very \textquotedblleft strong\textquotedblright}\\
\text{\textit{completely}}%
\end{array}
\end{array}
&
\begin{array}
[c]{c}%
\mathbf{d}_{\mathbf{G}}=\mathbf{3}\\%
\begin{array}
[c]{c}%
\text{\textquotedblleft strong\textquotedblright}\\
\text{\textit{in 2 cases}}%
\end{array}
\end{array}
&
\begin{array}
[c]{c}%
\mathbf{d}_{\mathbf{T/U}}=\mathbf{2}\\%
\begin{array}
[c]{c}%
\text{\textquotedblleft weak\textquotedblright}\\
\text{\textit{in 2 cases}}%
\end{array}
\end{array}
&
\begin{array}
[c]{c}%
\mathbf{d}_{\mathbf{A}}=\mathbf{1}\\%
\begin{array}
[c]{c}%
\text{very \textquotedblleft weak\textquotedblright}\\
\text{\textit{never}}%
\end{array}
\end{array}
\end{array}
\label{1}%
\end{equation}
which allows us to make transition from qualitative to quanti\-ta\-tive
description of genetic code structure \cite{e-dup/dup1,dup/dup3}.

We use the notation $\mathbf{T/U}$, because genetic code is read from mRNA,
and so we will not differentiate their determinative ability
(\textquotedblleft strength\textquotedblright) in what follows.

Let us present four bases (\ref{1}) as the vector-column%

\begin{equation}
\mathbb{V}=\left(
\begin{array}
[c]{c}%
\mathbf{V}_{1}\\
\mathbf{V}_{2}\\
\mathbf{V}_{3}\\
\mathbf{V}_{4}%
\end{array}
\right)  =\left(
\begin{array}
[c]{c}%
\mathbf{C}^{\left(  4\right)  }\\
\mathbf{G}^{\left(  3\right)  }\\
\mathbf{T}^{\left(  2\right)  }\\
\mathbf{A}^{\left(  1\right)  }%
\end{array}
\right)  \label{v}%
\end{equation}
and the corresponding the vector-row
\begin{equation}
\mathbb{V}^{T}=\left(
\begin{array}
[c]{cccc}%
\mathbf{C}^{\left(  4\right)  } & \mathbf{G}^{\left(  3\right)  } &
\mathbf{T}^{\left(  2\right)  } & \mathbf{A}^{\left(  1\right)  }%
\end{array}
\right)  . \label{vt}%
\end{equation}
where the upper index for nucleotide denotes determinative degree. We make the
exterior product of vector-column (\ref{v}) and vector-row (\ref{vt}) as
follows \cite{e-dup/dup1,e-dup/dup0}
\begin{equation}
\mathbb{M}=\mathbb{V}\times\mathbb{V}^{T}=\left(
\begin{array}
[c]{cccc}%
\mathbf{C}^{(4)}\mathbf{C}^{(4)} & \mathbf{C}^{(4)}\mathbf{G}^{\left(
3\right)  } & \mathbf{C}^{(4)}\mathbf{T}^{\left(  2\right)  } & \mathbf{C}%
^{(4)}\mathbf{A}^{\left(  1\right)  }\\
\mathbf{G}^{\left(  3\right)  }\mathbf{C}^{(4)} & \mathbf{G}^{\left(
3\right)  }\mathbf{G}^{\left(  3\right)  } & \mathbf{G}^{\left(  3\right)
}\mathbf{T}^{\left(  2\right)  } & \mathbf{G}^{\left(  3\right)  }%
\mathbf{A}^{\left(  1\right)  }\\
\mathbf{T}^{\left(  2\right)  }\mathbf{C}^{(4)} & \mathbf{T}^{\left(
2\right)  }\mathbf{G}^{\left(  3\right)  } & \mathbf{T}^{\left(  2\right)
}\mathbf{T}^{\left(  2\right)  } & \mathbf{T}^{\left(  2\right)  }%
\mathbf{A}^{\left(  1\right)  }\\
\mathbf{A}^{\left(  1\right)  }\mathbf{C}^{(4)} & \mathbf{A}^{\left(
1\right)  }\mathbf{G}^{\left(  3\right)  } & \mathbf{A}^{\left(  1\right)
}\mathbf{T}^{\left(  2\right)  } & \mathbf{A}^{\left(  1\right)  }%
\mathbf{A}^{\left(  1\right)  }%
\end{array}
\right)  . \label{vv}%
\end{equation}

It is remarkable that the matrix $\mathbb{M}$ (\ref{vv}) fully
\textit{coincides} with the canonical matrix of doublets (\ref{0}), if and
only if the vector $\mathbb{V}$ has the determinative degree order
$\mathbf{C}$\textbf{, }$\mathbf{G}$\textbf{, }$\mathbf{U}$\textbf{,
}$\mathbf{A}$ (\ref{1}). Although there are 4!=24 possibilities to place 4
bases in row, but all others except one presented in (\ref{1}) do not reflect
phenomenological properties of genetic code. It follows that the intuitive
\textquotedblleft rhombic code\textquotedblright\ and genetic
vocabulary\ \cite{e-rum1,e-kar/sor,e-kara1} have their own inner abstract
structure uniquely defined by exterior product of special vectors (\ref{v}).
This ordering is also adjusted to the schemes \cite{e-sukh,e-mas}, also
(partially) with half time of nucleotide substitution under mutational
pressure \cite{kow/mac/ceb} and the nucleotides information weights
\cite{dud/ceb/kow/mac/now}. Indeed these facts allows us to introduce the
determinative degree, as an \textit{abstract variable} being a numerical
measure of nucleotide difference in ability to determine sense of codon
\cite{e-dup/dup1,e-dup/dup0}.

Analogous model for the triplet genetic code can be constructed using triple
exterior product in the same way \cite{e-dup/dup1}. We dispose the doublet
matrix $\mathbb{M}$ on the \textbf{XY} plane and multiply it on the
vector-column $\mathbb{V}$ (\ref{v}) disposed along \textbf{Z} axis, i.e. we
construct the triple exterior product
\begin{equation}
\mathbb{K}=\mathbb{V}\times\mathbb{M}. \label{k}%
\end{equation}

Thus we obtain three-dimensional matrix over set of all triplets, and, since
each codon (except three terminal ones) corresponds to an amino acid, that can
be treated as a \textit{cubic matrix model of the genetic code}
\cite{e-dup/dup1}.

\section{Determinative degree and nucleotide properties}

The connection bulk DNA structure and various properties of nucleotides was
studies in \cite{zhe/sam/gov,gov/dan/mis/kon}. It is well-known that by
chemical structure the 4 nitrous bases can be divided into:

1) purine (\textbf{A},\textbf{G}) and pyrimidine (\textbf{C},\textbf{T});

2) having amino (\textbf{A},\textbf{C}) group and (\textbf{G},\textbf{T}) keto group;

3) making 3 (strong) hydrogen bonds (\textbf{C},\textbf{G}) and 2 (weak)
hydrogen bonds (\textbf{A},\textbf{T}).

They give rise to 3 symmetry transformations:

1) Purine-pyrimidine symmetry%
\begin{equation}
\left(
\begin{array}
[c]{c}%
\mathbf{T}^{\left(  2\right)  }\\
\mathbf{A}^{\left(  1\right)  }\\
\mathbf{C}^{\left(  4\right)  }\\
\mathbf{G}^{\left(  3\right)  }%
\end{array}
\right)  =\left(
\begin{array}
[c]{cccc}%
0 & 0 & 1 & 0\\
0 & 0 & 0 & 1\\
1 & 0 & 0 & 0\\
0 & 1 & 0 & 0
\end{array}
\right)  \left(
\begin{array}
[c]{c}%
\mathbf{C}^{\left(  4\right)  }\\
\mathbf{G}^{\left(  3\right)  }\\
\mathbf{T}^{\left(  2\right)  }\\
\mathbf{A}^{\left(  1\right)  }%
\end{array}
\right)  =\mathcal{R}_{pur}\mathbb{V;} \label{sym-p}%
\end{equation}

2) Amino-keto symmetry%

\begin{equation}
\left(
\begin{array}
[c]{c}%
\mathbf{A}^{\left(  1\right)  }\\
\mathbf{T}^{\left(  2\right)  }\\
\mathbf{G}^{\left(  3\right)  }\\
\mathbf{C}^{\left(  4\right)  }%
\end{array}
\right)  =\left(
\begin{array}
[c]{cccc}%
0 & 0 & 0 & 1\\
0 & 0 & 1 & 0\\
0 & 1 & 0 & 0\\
1 & 0 & 0 & 0
\end{array}
\right)  \left(
\begin{array}
[c]{c}%
\mathbf{C}^{\left(  4\right)  }\\
\mathbf{G}^{\left(  3\right)  }\\
\mathbf{T}^{\left(  2\right)  }\\
\mathbf{A}^{\left(  1\right)  }%
\end{array}
\right)  =\mathcal{R}_{amino}\mathbb{V;} \label{sym-a}%
\end{equation}

3) Complementary symmetry (leaving the double helix invariant)%
\begin{equation}
\left(
\begin{array}
[c]{c}%
\mathbf{G}^{\left(  3\right)  }\\
\mathbf{C}^{\left(  4\right)  }\\
\mathbf{A}^{\left(  1\right)  }\\
\mathbf{T}^{\left(  2\right)  }%
\end{array}
\right)  =\left(
\begin{array}
[c]{cccc}%
0 & 1 & 0 & 0\\
1 & 0 & 0 & 0\\
0 & 0 & 0 & 1\\
0 & 0 & 1 & 0
\end{array}
\right)  \left(
\begin{array}
[c]{c}%
\mathbf{C}^{\left(  4\right)  }\\
\mathbf{G}^{\left(  3\right)  }\\
\mathbf{T}^{\left(  2\right)  }\\
\mathbf{A}^{\left(  1\right)  }%
\end{array}
\right)  =\mathcal{R}_{compl}\mathbb{V,} \label{sym-c}%
\end{equation}
where the even (because determinant is +1) permutation matrices $\mathcal{R}%
_{pur},\mathcal{R}_{amino},\mathcal{R}_{compl}$ satisfy%
\[
\mathcal{R}_{pur}\mathcal{R}_{amino}\mathcal{R}_{compl}=\mathcal{I},
\]
and two of them, e.g. $\mathcal{R}_{pur}$, $\mathcal{R}_{compl}$ together with
the identity matrix $\mathcal{I}$ form the dihedral group $D_{2}$ which is the
symmetry group of the dihedron, or regular double-pyramid, with vertices on
the unit-sphere (see e.g. \cite{ziegler}). Another representation of this
group by $3\times3$ rotational matrices is called a DNA\ group \cite{zha1}.

The difference in the number of hydrogen bonds causes the different
interaction with its complementary nucleotide: each \textquotedblleft
strong\textquotedblright\ nucleotide ($\mathbf{C}$ and $\mathbf{G}$) has 3
bonds and the energy of \textbf{C}-\textbf{G} interaction is -2.4 kkal/mol,
and each \textquotedblleft weak\textquotedblright\ nucleotide ($\mathbf{T}$
and $\mathbf{A}$) has only 2 bonds and the energy of \textbf{A}-\textbf{T}
interaction is -1.2 kkal/mol \cite{e-lewi}. Therefore each base has its own
properties and so dividing them into only 2 groups is not sufficient.

We then can search whether the ordering (\ref{1}) is adjusted to some physical
properties of nucleotides.

\begin{figure}[tbh]
\centerline{\includegraphics{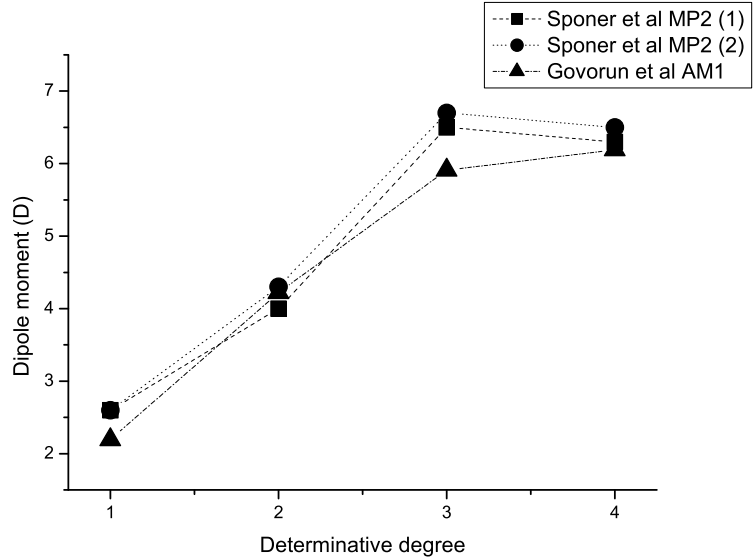}}\caption{Dipole moment of DNA\ bases
calculated by methods AM1 \cite{gov/dan/mis/kon} (triangles) and two
modifications of MP2 \cite{spo/les/vet/hob} (squares and circles). The
corresponding linear fits are: $D_{\mathrm{AM1}}=1.21+1.37\mathbf{d}_{x}$
($R=0.96)$; $D_{\mathrm{MP2(1)}}=1.45+1.36\mathbf{d}_{x}$ ($R=0.93)$;
$D_{\mathrm{MP2(2)}}=1.5+1.41\mathbf{d}_{x}$ ($R=0.93)$. }%
\label{f-dip}%
\end{figure}

First we observe that the dipole moment of bases is proportional to the
determinative degree as it is shown on Fig. \ref{f-dip}.

\begin{figure}[tbh]
\centerline{\includegraphics{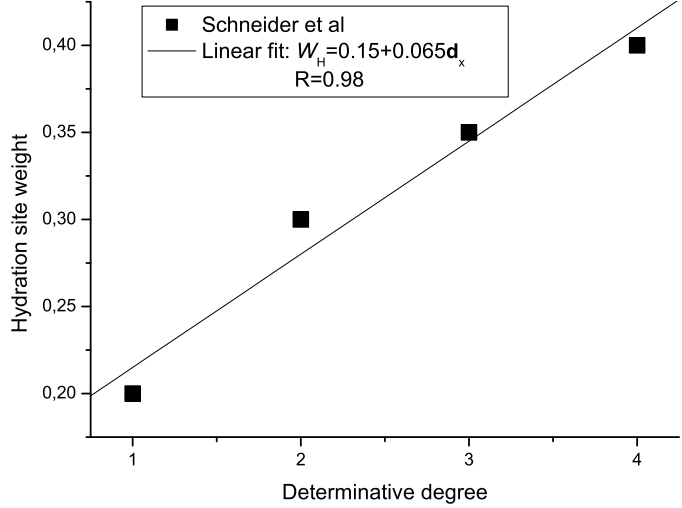}}\caption{Weights of hydration site
\cite{sch/ber}.}%
\label{f-hyd}%
\end{figure}

Then we see that the weight of hydration sites for bases is also proportional
to the determinative degree Fig. \ref{f-hyd}.

We can conclude that the determinative degree reflects not only redundancy of
genetic code in the third position, but also connected with some energetic
properties of bases themselves.

\section{Trianders and their characteristics}

It can be assumed that the phenomenological properties of genetic code and
inequality of bases (reflected in (\ref{1})) will become apparent in real
nucleotide sequences. Here we use the introduced determinative degree to build
a new kind of sequence analysis based on some special modification of
DNA\ walks method \cite{azb,lobr1,ceb/dud,ber/gla/sko}.

\subsection{Triander construction}

We embed a nucleotide sequence into the two-dimensional \textit{determinative
degree space} (DD plane) in the following way. The axis assignment corresponds
to the value of nucleotide determinative degree as%
\begin{align*}
\text{Axis }x  &  \text{:}\text{ }\left\{  \mathbf{A}\right\}  =\left(
-1,0\right)  ;\left\{  \mathbf{T}\right\}  =\left(  +2,0\right)  ,\\
\text{Axis }y  &  \text{:}\text{ }\left\{  \mathbf{G}\right\}  =\left(
0,-3\right)  ;\left\{  \mathbf{C}\right\}  =\left(  0,+4\right)  .
\end{align*}

Moving along a sequence produces a walk in the determinative degree space
which we call a \textit{determinative degree walk}. In general, a current
point on DD plane after $i$ steps is determined by the coordinates%
\begin{align}
x_{i}^{DD}  &  =\mathbf{d}_{\mathbf{T}}n_{\mathbf{T}}\left(  i\right)
-\mathbf{d}_{\mathbf{A}}n_{\mathbf{A}}\left(  i\right)  ,\label{x}\\
y_{i}^{DD}  &  =\mathbf{d}_{\mathbf{C}}n_{\mathbf{C}}\left(  i\right)
-\mathbf{d}_{\mathbf{G}}n_{\mathbf{G}}\left(  i\right)  , \label{y}%
\end{align}
where $n_{\mathbf{X}}\left(  i\right)  $ is cumulative quantity of nucleotide
$\mathbf{X}$ after $i$ steps and $\mathbf{d}_{\mathbf{X}}$ is the
determinative degree of nucleotide $\mathbf{X}$. The standard DNA walks
\cite{ber/gla/sko} (genome landscapes \cite{lobr1}) have all $\mathbf{d}%
_{\mathbf{X}}=1$ in (\ref{x})--(\ref{y}), i.e.%
\begin{align}
x_{i}^{standard}  &  =n_{\mathbf{T}}\left(  i\right)  -n_{\mathbf{A}}\left(
i\right)  ,\label{x0}\\
y_{i}^{standard}  &  =n_{\mathbf{C}}\left(  i\right)  -n_{\mathbf{G}}\left(
i\right)  . \label{y0}%
\end{align}
The one-dimensional (purine/pyrimidine) DNA walks are defined by only one
coordinate, while $x$ is chosen as position, i.e.%
\begin{align}
x_{i}^{pp}  &  =i,\label{xp}\\
y_{i}^{pp}  &  =n_{\mathbf{C}}\left(  i\right)  +n_{\mathbf{T}}\left(
i\right)  -n_{\mathbf{A}}\left(  i\right)  -n_{\mathbf{G}}\left(  i\right)  .
\label{yp}%
\end{align}

Therefore, while \textquotedblleft purine/pyrimidine\textquotedblright\ DNA
walks manifestly show the purine/pyrimidine imbalance, the standard DNA walks
(\ref{x0})--(\ref{y0}) applied for one strand show DNA asymmetry
\cite{wu,fra/och} (violation of the Parity Rule 2 \cite{sue1}), the
determinative degree walk (\ref{x})--(\ref{y}) visually shows
\textquotedblleft strength\textquotedblright\ imbalance in one strand.

Then we build 3 independent determinative degree walks beginning from 1st
nucleotide with step 3 (due to the triplet structure of genetic code). In this
way we obtain 3 broken lines (\textit{branches}) starting from the point of
origin, and each of them presents the determinative degree walk through the
following nucleotide numbers:

\textbf{1st} branch goes through \textbf{1,4,7,10,13}... positions;

\textbf{2nd} branch goes through \textbf{2,5,8,11,14}... positions;

\textbf{3rd} branch goes through \textbf{3,6,9,12,15}... positions.

These 3 branches on the determinative degree plane are called
\textit{triander}.

If 1st letter corresponds to the first start codon nucleotide, then the
triander branches represent nucleotide sets in three codon positions indepen\-dently.

As distinct from previous versions of DNA\ walks in which all 4 nucleotides
are regarded equivalent in the sense they give equal by module shifts, in our
approach each nucleotide gives contribution different by module (which is
taken equal to its determinative degree). So, despite we obtain at first sight
isomorphic to \cite{ceb/dud} plot, trianders show not only quantitative
composition and pure statistical laws of symbol strings, but also
\textit{reflect} \textit{connection} between nucleotide sequences and inner
phenomenological properties of genetic code and physicochemical properties of bases.

\begin{figure}[tbh]
\centerline{\includegraphics[height=8cm,angle=-90]{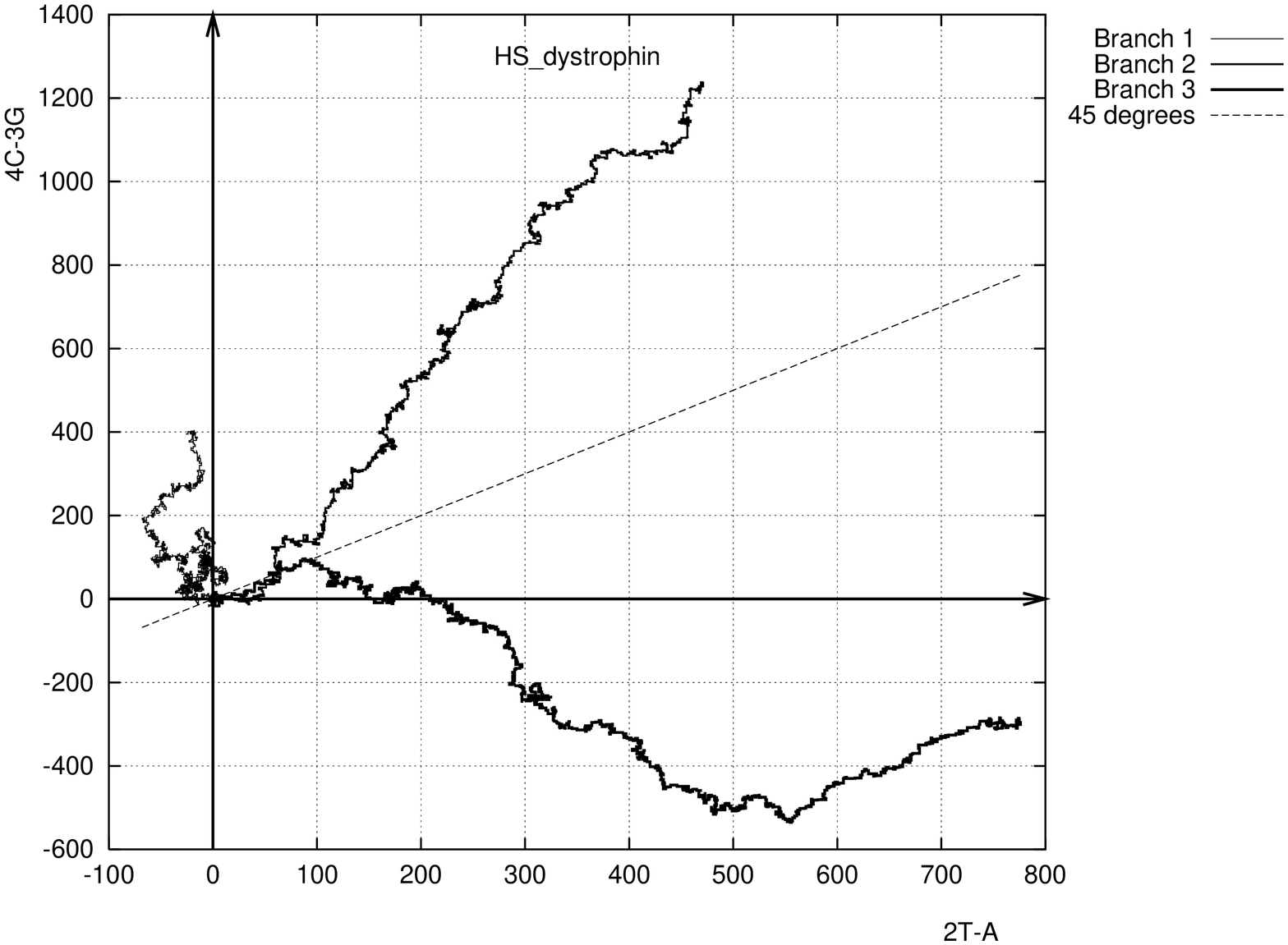}
\includegraphics[height=8cm,angle=-90]{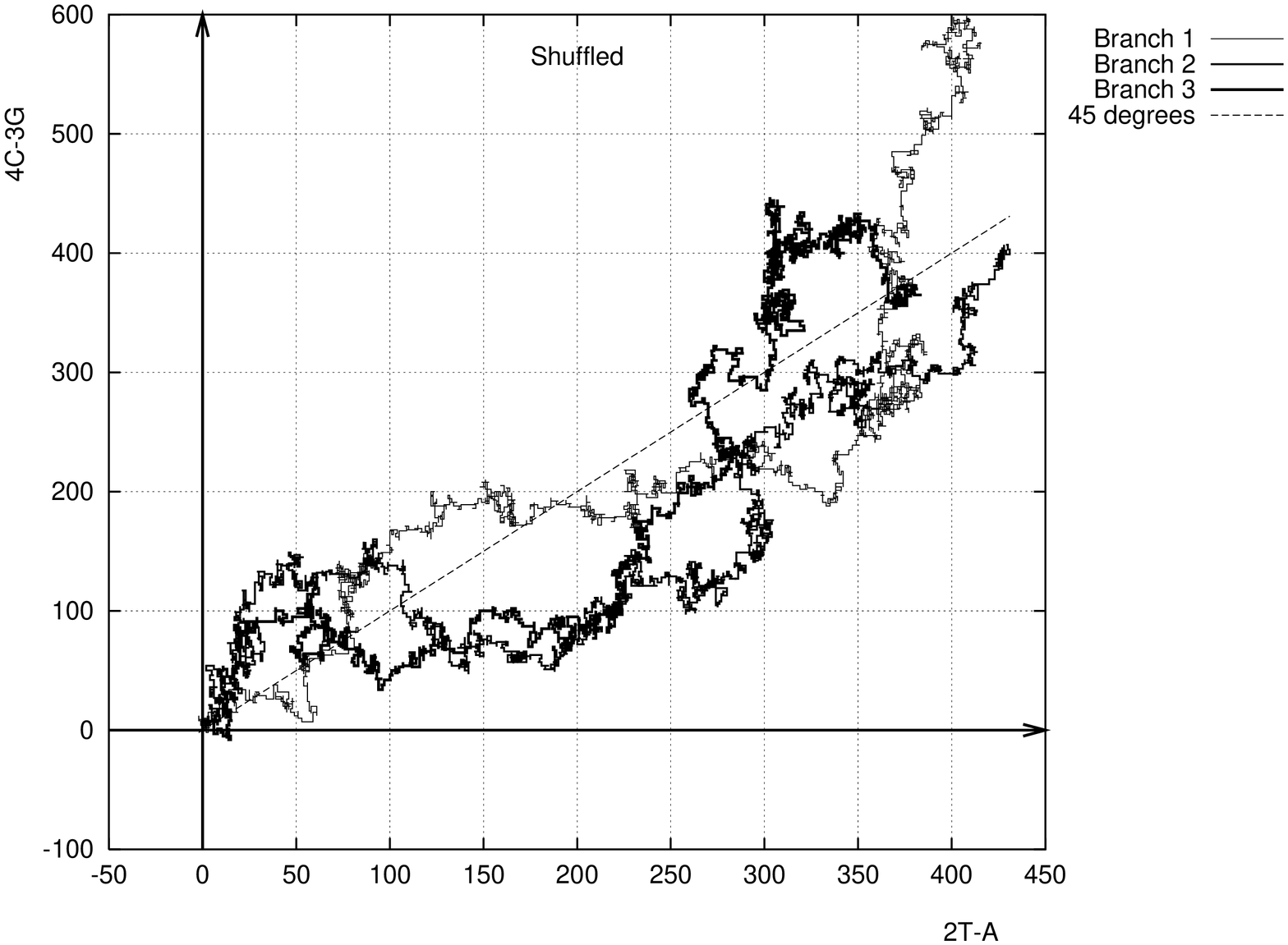}}\caption{Triander for the
Homo Sapiens dystrophin gene (left) and a shuffled sequence of the same
nucleotide composition (right).}%
\label{f-tri}%
\end{figure}

As an example of triander we will take the dystrophin gene which is the
largest gene found in nature, measuring 2.4 Mb, and is responsible for
Duchenne (DMD) and Becker (BMD) muscular dystrophies
\cite{yag/tak/wad/nak/mat}. The dystrophin RNA is differentially spliced,
producing a range of different transcripts, encoding a large set of protein
isoforms. Dystrophin is a large, rod-like cytoskeletal protein which is found
at the inner surface of muscle fibers. The triander for the dystrophin gene is
presented on Fig. \ref{f-tri}. For comparison we also show the triander for a
shuffled sequence of the same nucleotide composition. Obviously the ideal
triander for uniformly random sequence consists of 3 flowing together lines
from the origin having 45 degrees slope. This line also corresponds to the
symmetric sequence satisfying the Parity Rule 2 \cite{sue1}:
$N_{\mathbf{C}}=N_{\mathbf{G}}$, $N_{\mathbf{T}}=N_{\mathbf{A}}$. Such lines
are presented on all triander plots below for normalization.

\subsection{Determinative degree angle}

An important visual characteristic of a triander is the slop of its branches,
we call it determinative degree (DD) angle, which for a current point can be
calculated by%
\[
\tan\alpha\left(  i\right)  =\dfrac{4n_{\mathbf{C}}\left(  i\right)
-3n_{\mathbf{G}}\left(  i\right)  }{2n_{\mathbf{T}}\left(  i\right)
-n_{\mathbf{A}}\left(  i\right)  }.
\]

\begin{figure}[th]
\centerline{\includegraphics[height=8cm,angle=-90]{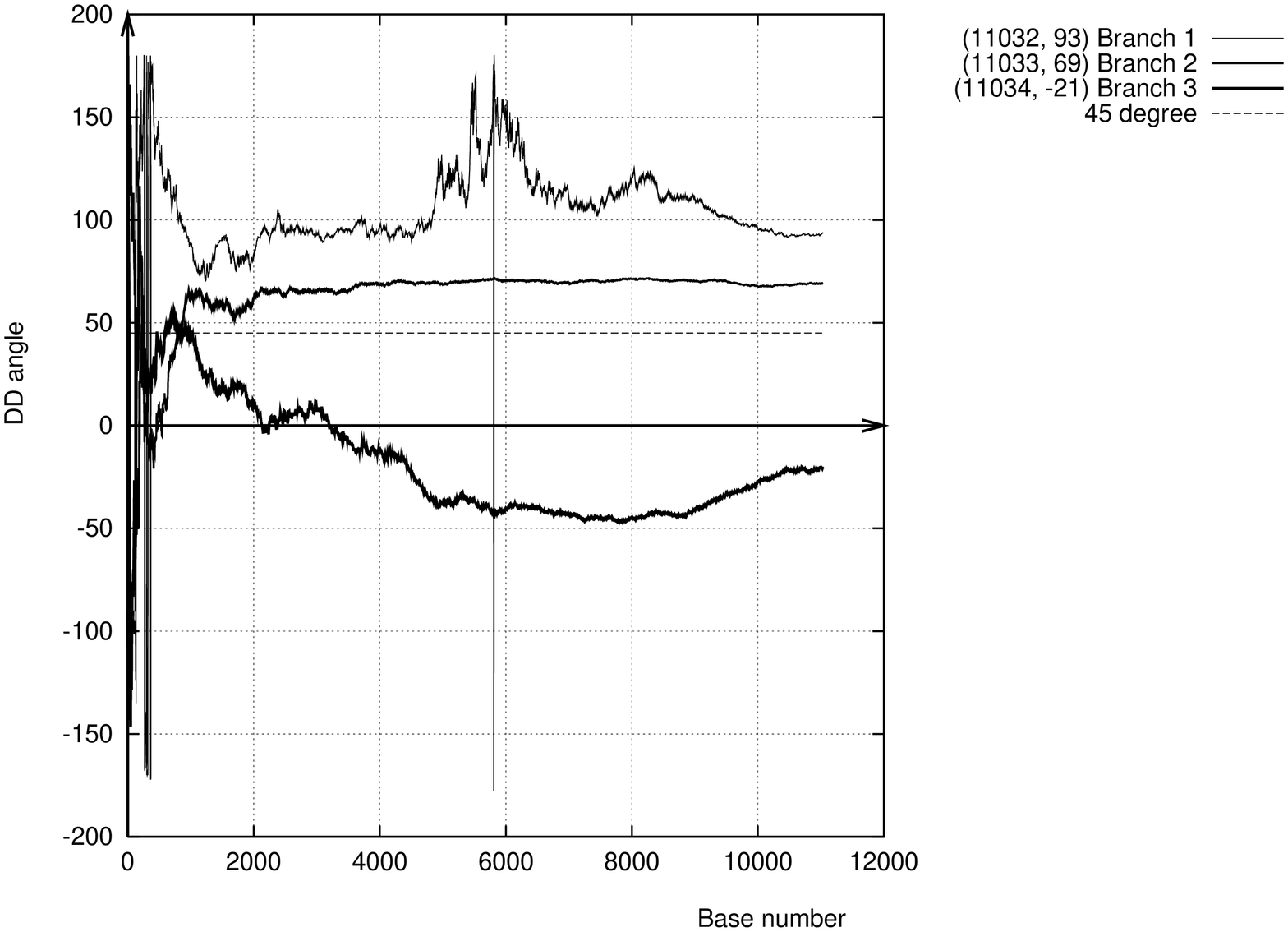}
\includegraphics[height=8cm,angle=-90]{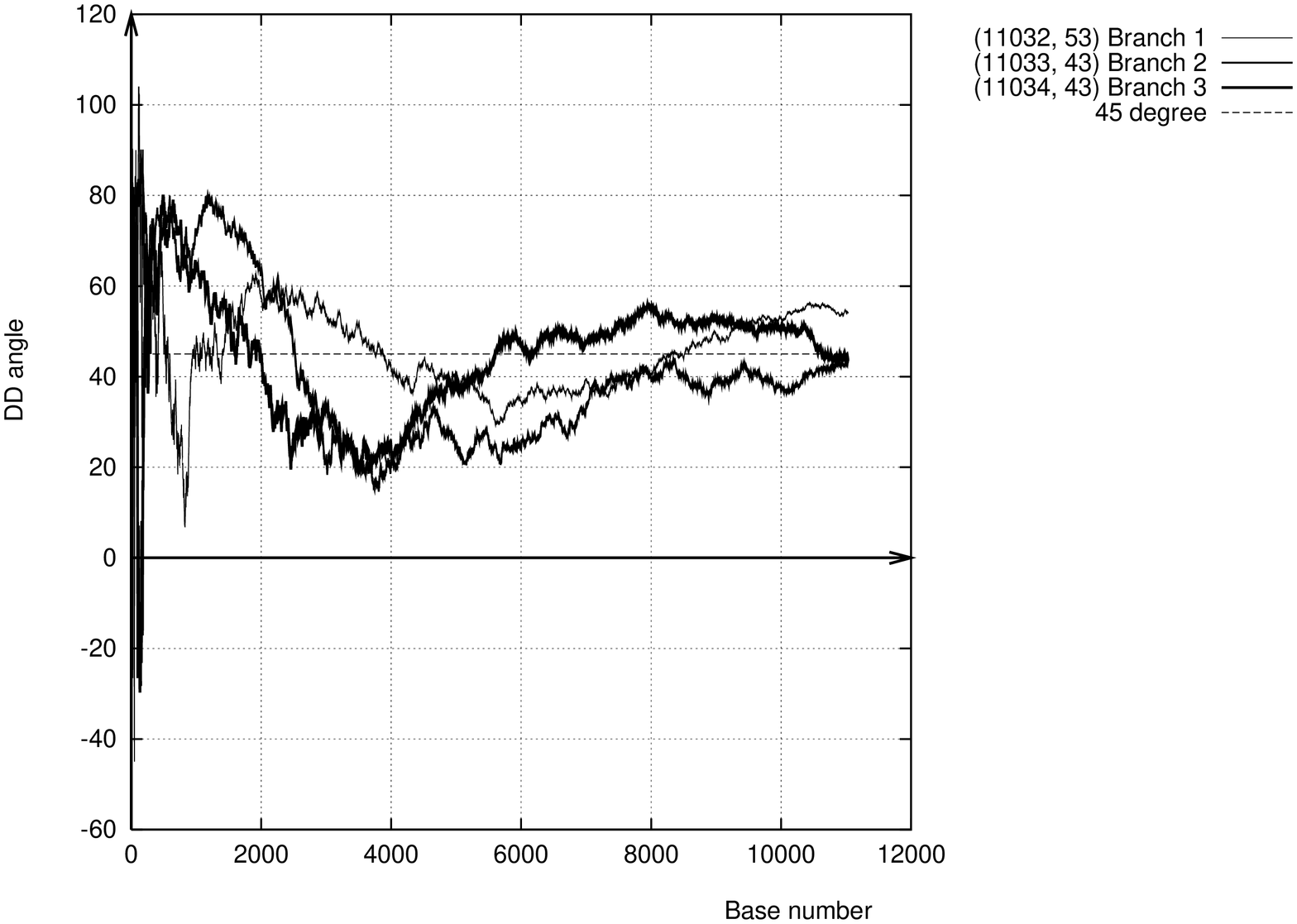}}\caption{Current
DD angle for triander of the Homo Sapiens dystrophin gene (left) and a
shuffled sequence of the same nucleotide composition (right).}%
\label{f-ang}%
\end{figure}

Here and below $n_{\mathbf{X}}$ denotes cumulative quantity of nucleotide $\mathbf{X}$
for a given branch.
Evidently, for uniformly random sequence or a symmetric sequence satisfying
the Parity Rule 2 \cite{sue1} the angle will be 45 degree (horizontal dashed
line of the below plots), and so the difference from this value will say about
nontrivial ordering. The plots of current values of $\alpha$ for the
dystrophin gene and for a shuffled sequence of the same nucleotide composition
are presented on Fig. \ref{f-ang}.

We stress that trianders show not only quantitative composition, but allow us
to find local motives in a more clear way, because different modules for
nucleotides lead to less number of superposition and selfintersections. Also
trianders more accurately reflect the tendency of the sequence as a whole
similarly to DNA walks. Thus triander can be treated as a \textquotedblleft
picture\textquotedblright, \textquotedblleft genome passport\textquotedblright%
\ or \textquotedblleft genogramma\textquotedblright\ of a given sequence.

If we remember that third base in codon has maximal redundancy, then 3rd
branch of a triander gets a definite \textquotedblleft physical
sense\textquotedblright. Let us assume that the determinative degree is an
additive variable (which can be made in first approximation at least
\cite{e-dup/dup1}), then 3rd branch can show current \textquotedblleft
strength\textquotedblright\ of sequence, that is the \textquotedblleft
bulk\textquotedblright\ ability to determine sense of codon. In this scheme
other two branches can be treated as 3rd branch with shifted ORF (Open Reading Frame).

\subsection{Euclidean and Manhattan distances}

As the measure of the \textquotedblleft sequence strength\textquotedblright%
\ we can choose length of the radius-vector from the origin to the current
point of triander, i.e. the Euclidean distance%
\[
D_{E}\left(  i\right)  =\sqrt{\left(  4n_{\mathbf{C}}\left(  i\right)
-3n_{\mathbf{G}}\left(  i\right)  \right)  ^{2}+\left(  2n_{\mathbf{T}}\left(
i\right)  -n_{\mathbf{A}}\left(  i\right)  \right)  ^{2}}.
\]

We can also use the Manhattan distance\footnote{Also known as
\textit{rectilinear distance,} and it can be treated as the distance that
would be traveled to get from one data point to the other if a grid-like path
is followed (a car driving in a city laid out in square blocks, like
Manhattan).}%
\[
D_{M}\left(  i\right)  =\left\vert 4n_{\mathbf{C}}\left(  i\right)
-3n_{\mathbf{G}}\left(  i\right)  \right\vert +\left\vert 2n_{\mathbf{T}%
}\left(  i\right)  -n_{\mathbf{A}}\left(  i\right)  \right\vert ,
\]
which is the distance between two points measured along axes at right angles
(see e.g. \cite{skiena}).

\begin{figure}[tbh]
\centerline{\includegraphics[height=8cm,angle=-90]{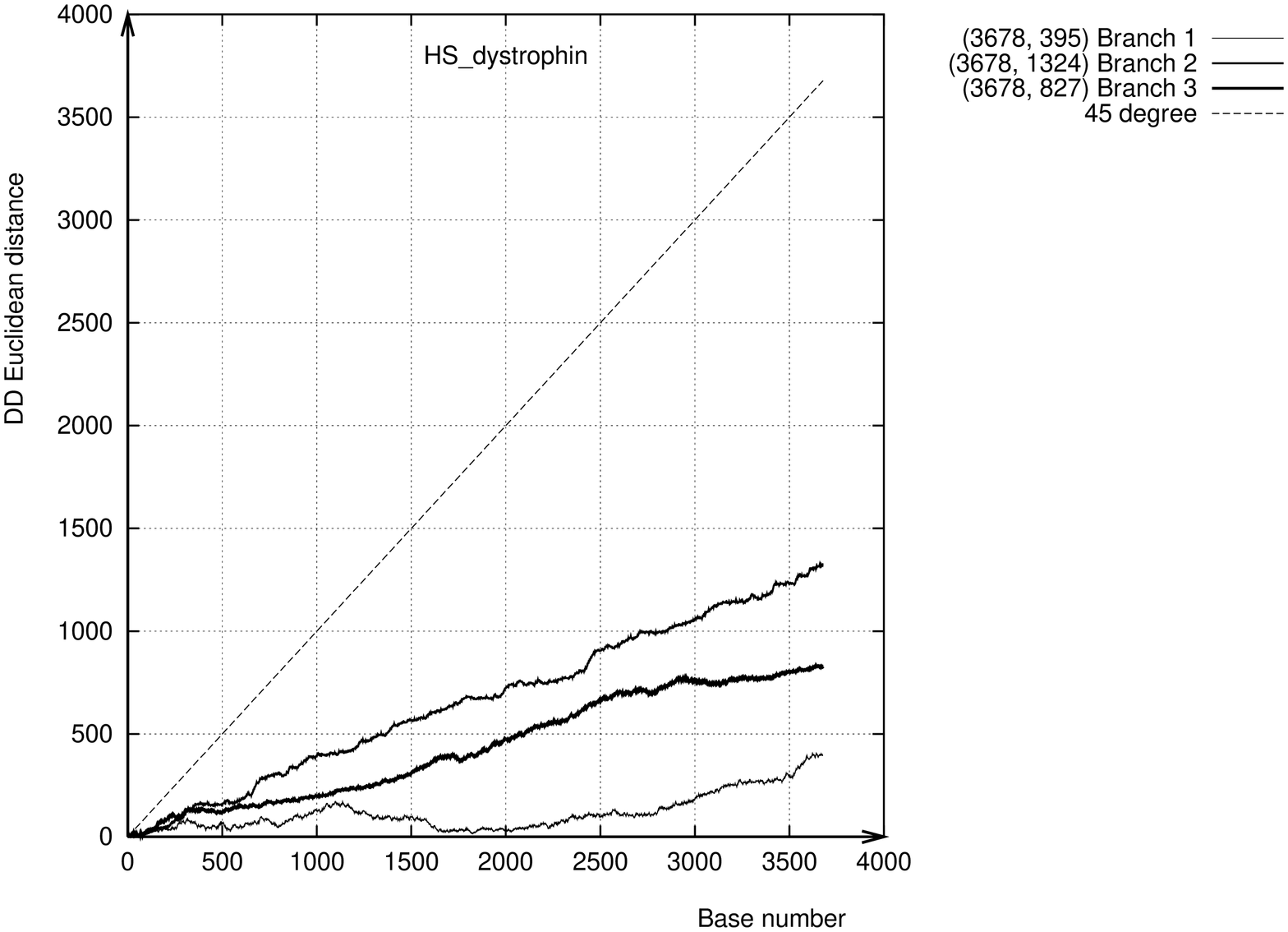}
\includegraphics[height=8cm,angle=-90]{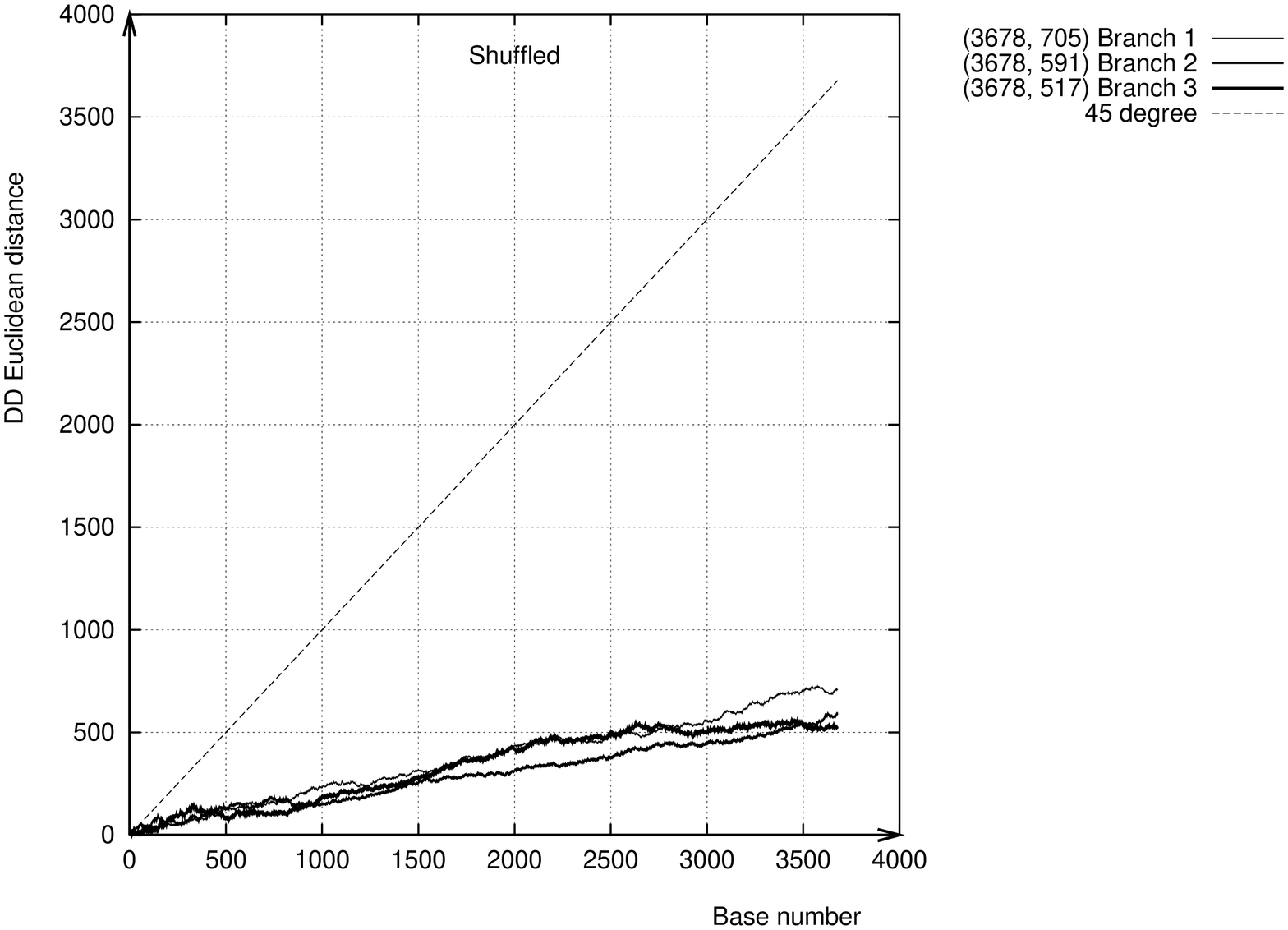}}\caption{Current
DD Euclidean distance for triander of the Homo Sapiens dystrophin gene (left)
and a shuffled sequence of the same nucleotide composition (right).}%
\label{f-eucl}%
\end{figure}

\begin{figure}[tbh]
\centerline{\includegraphics[height=8cm,angle=-90]{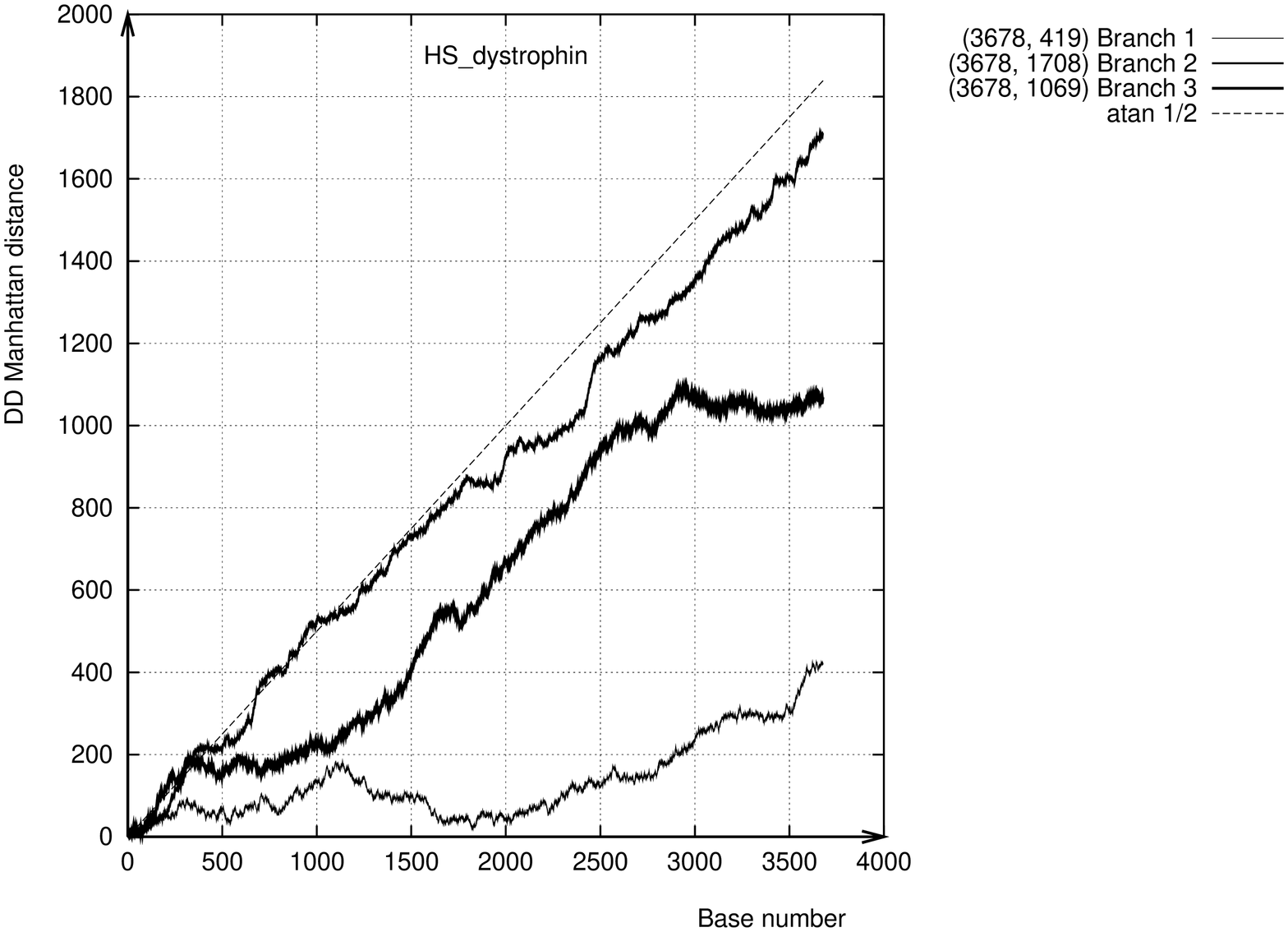}
\includegraphics[height=8cm,angle=-90]{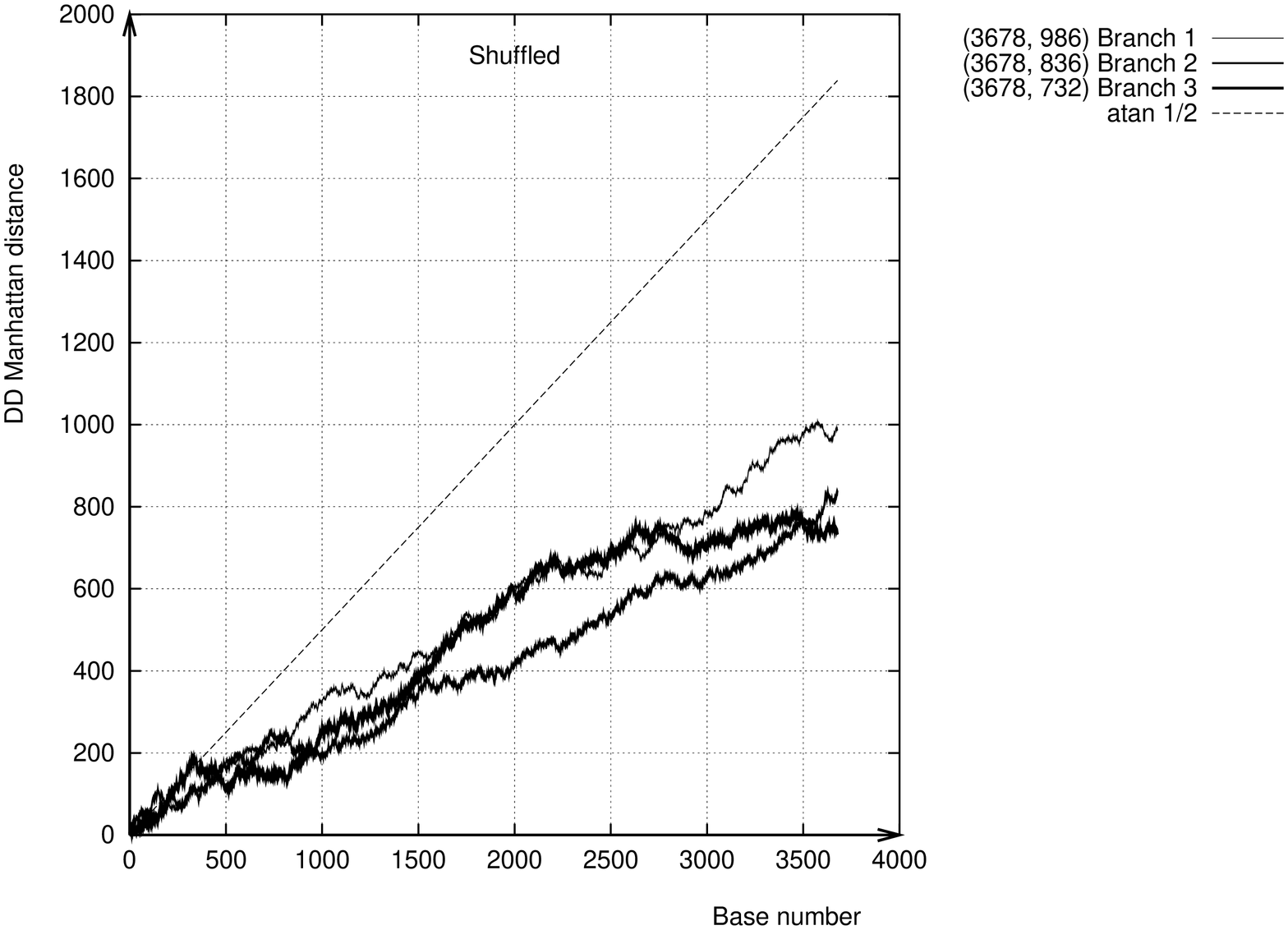}}\caption{Current
DD Manhattan distance for triander of the Homo Sapiens dystrophin gene (left)
and a shuffled sequence of the same nucleotide composition (right).}%
\label{f-mann}%
\end{figure}

In case of symmetric sequence (equal number of all nucleotides) at the step
$i$ the Euclidean and Manhattan distances are $D_{E}\left(  i\right)
=i/\sqrt{2}$ and $D_{M}\left(  i\right)  =i/2$ (which is shown by dashed lines
on Figs. \ref{f-eucl},\ref{f-mann}).

\subsection{Visualization of the genetic code triplet nature}

\begin{figure}[tbh]
\centerline{\includegraphics[height=8cm,angle=-90]{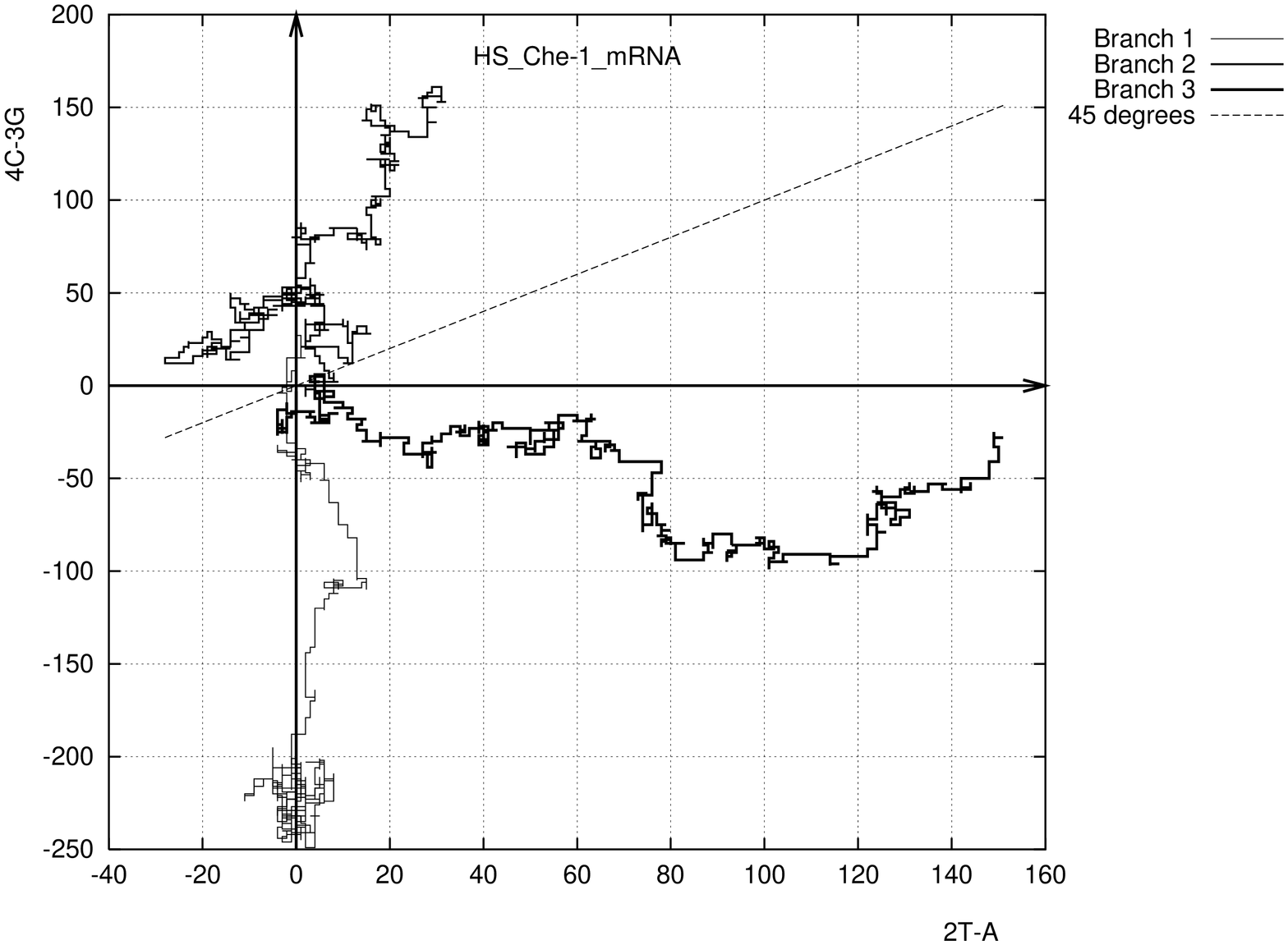}
\includegraphics[height=8cm,angle=-90]{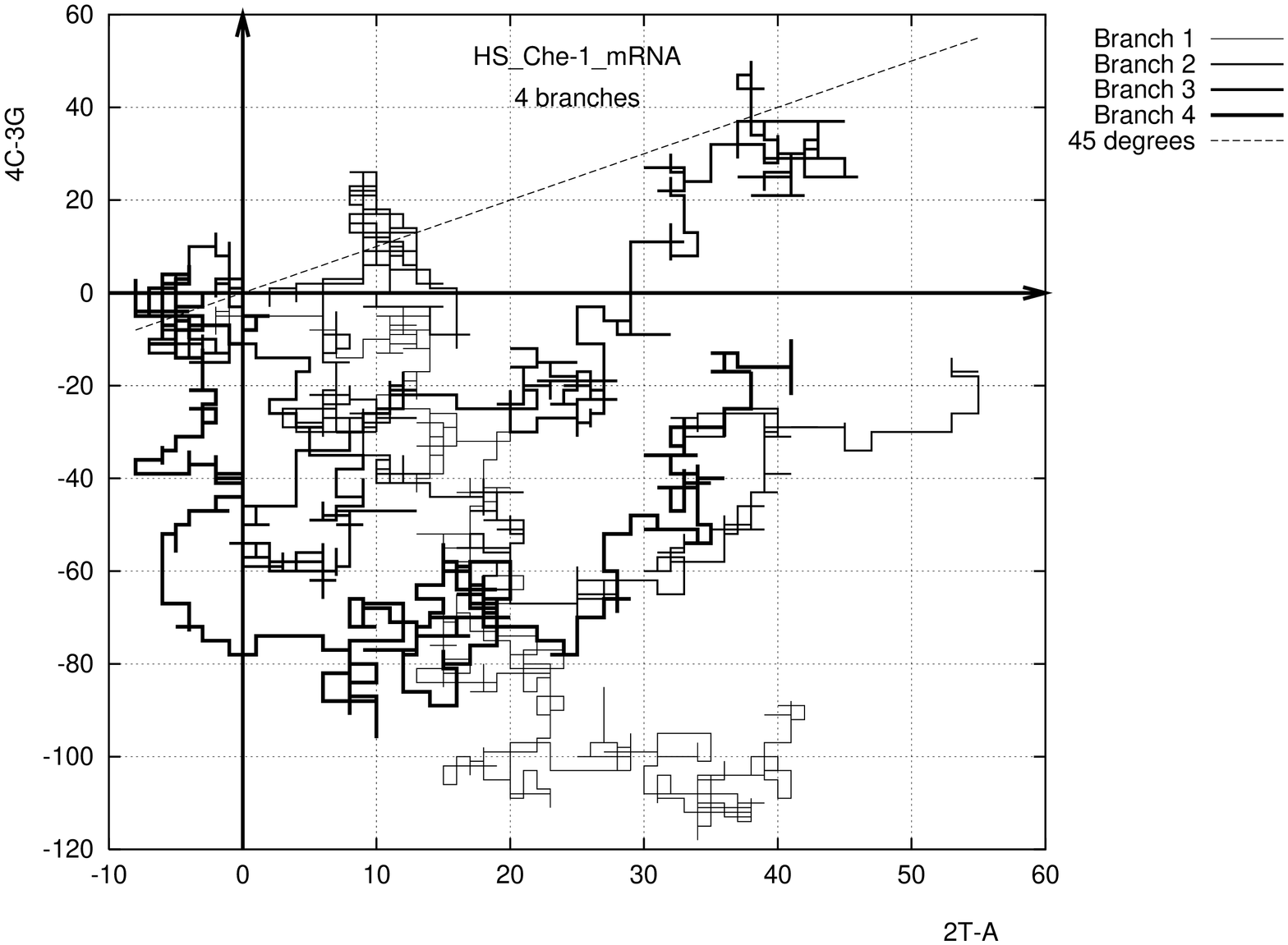}}
\centerline{\includegraphics[height=8cm,angle=-90]{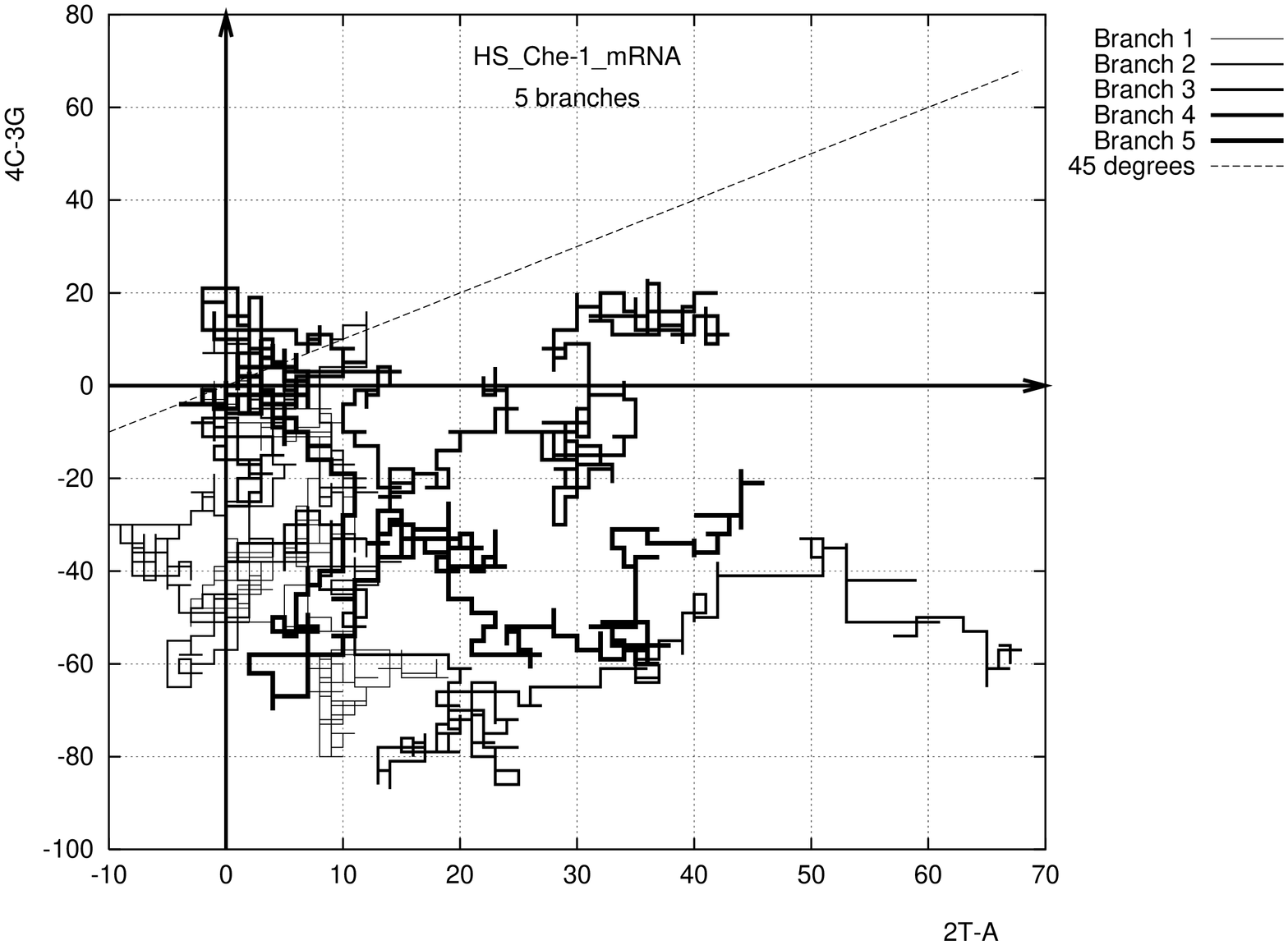}
\includegraphics[height=8cm,angle=-90]{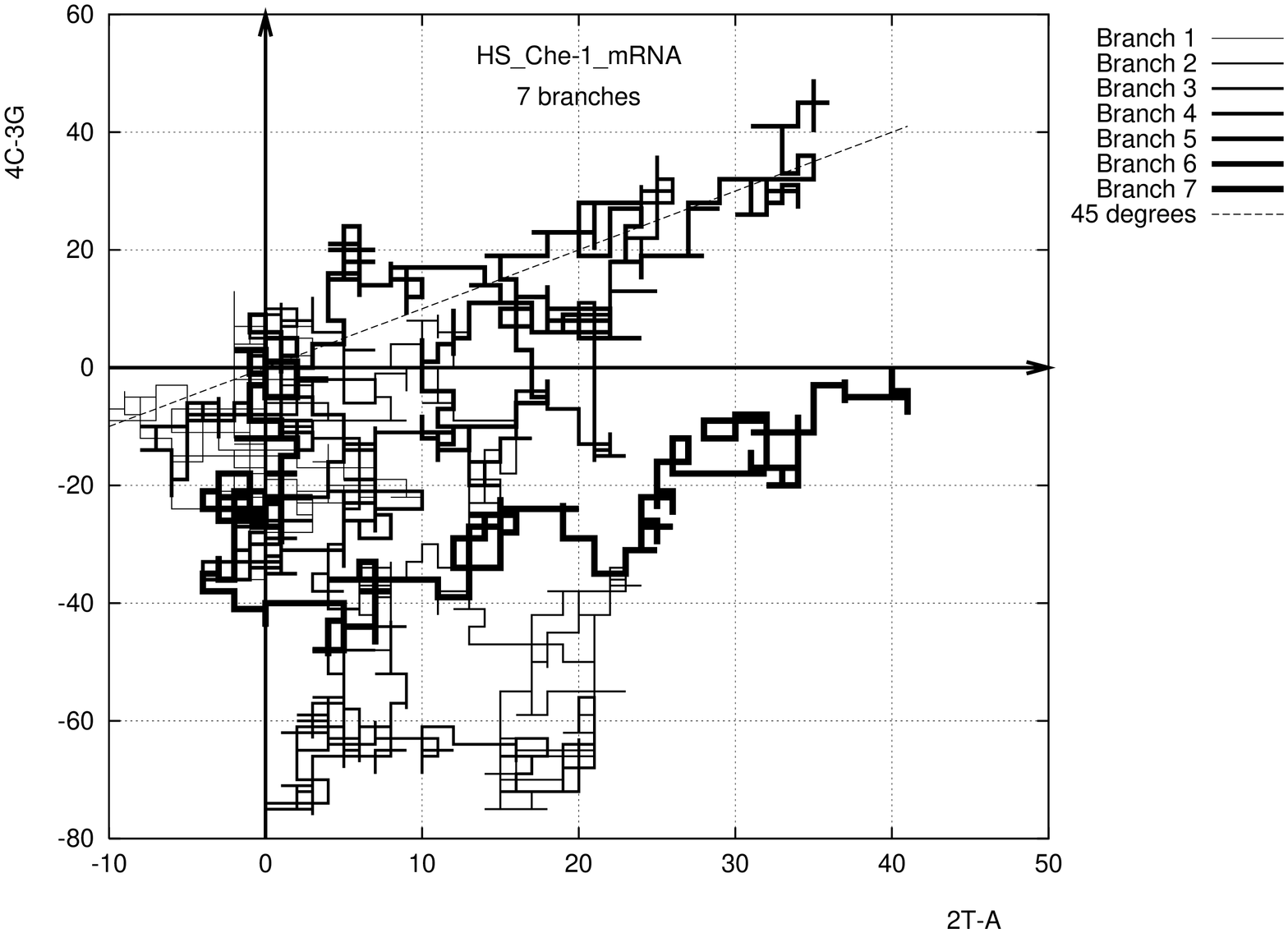}}\caption{Trianders of the
Homo sapiens Che-1 mRNA gene (first) and analogs of trianders with phases =
4,5,7 . The strong triplet signal is clearly seen.}%
\label{f-step}%
\end{figure}

Now we make sure that triplet character of the genetic code can be seen
directly from sequences representation by trianders. As an example we take
gene of Homo sapiens Che-1 mRNA. We consider additionally analogs of trianders
with different phases = 4,5,7. The result is presented on Fig. \ref{f-step},
from which it is seen that only the case phase = 3 provides nontrivial
ordering leading to definite branches, that is we have clear visual
presentation of the strong triplet signal.

In such a way one could search for higher phase statistical correlations and
possible structures, if any, in nucleotide sequences.

\subsection{Transformations of trianders}

\begin{figure}[tbh]
\centerline{\includegraphics[height=8cm,angle=-90]{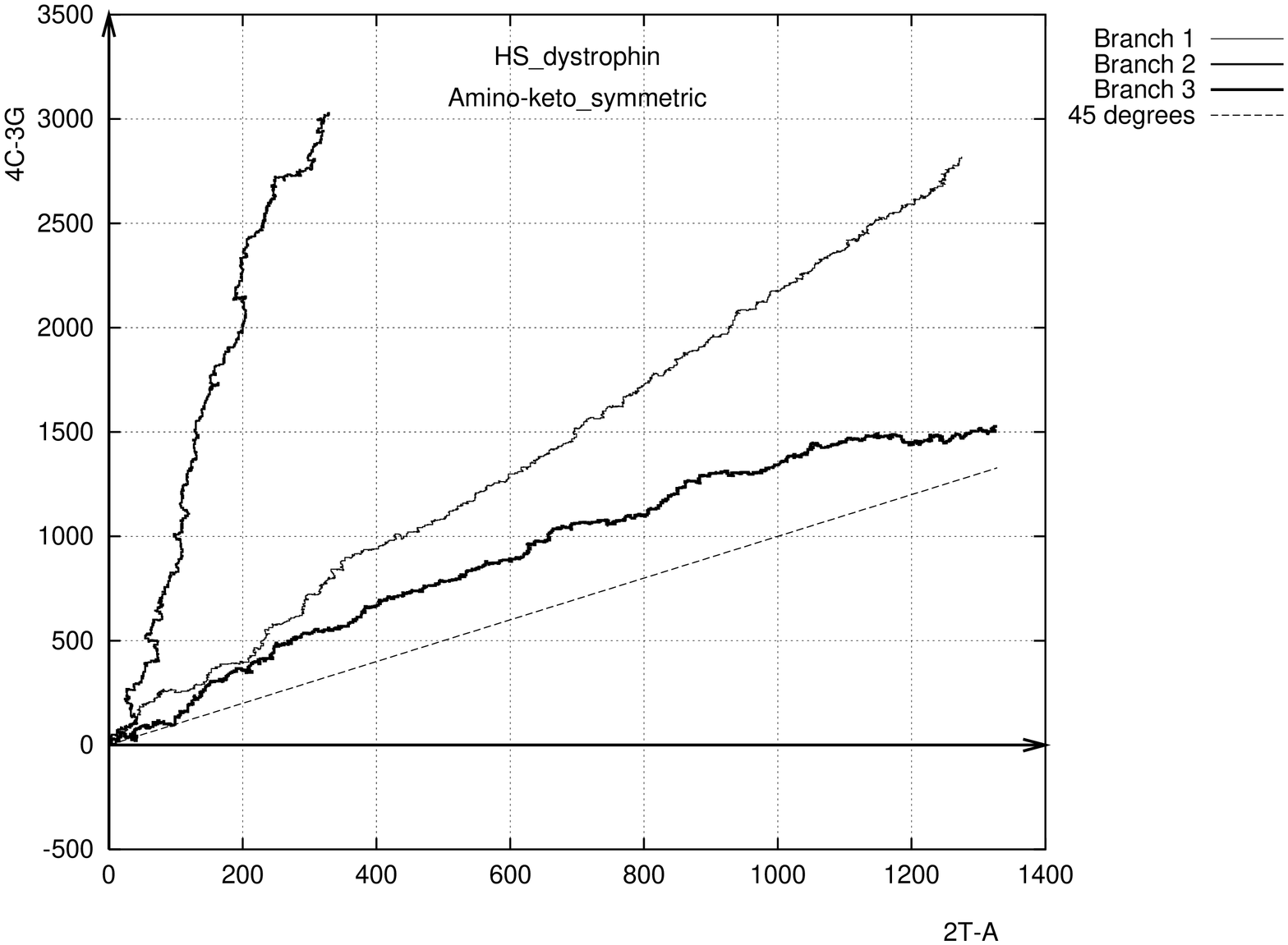}
\includegraphics[height=8cm,angle=-90]{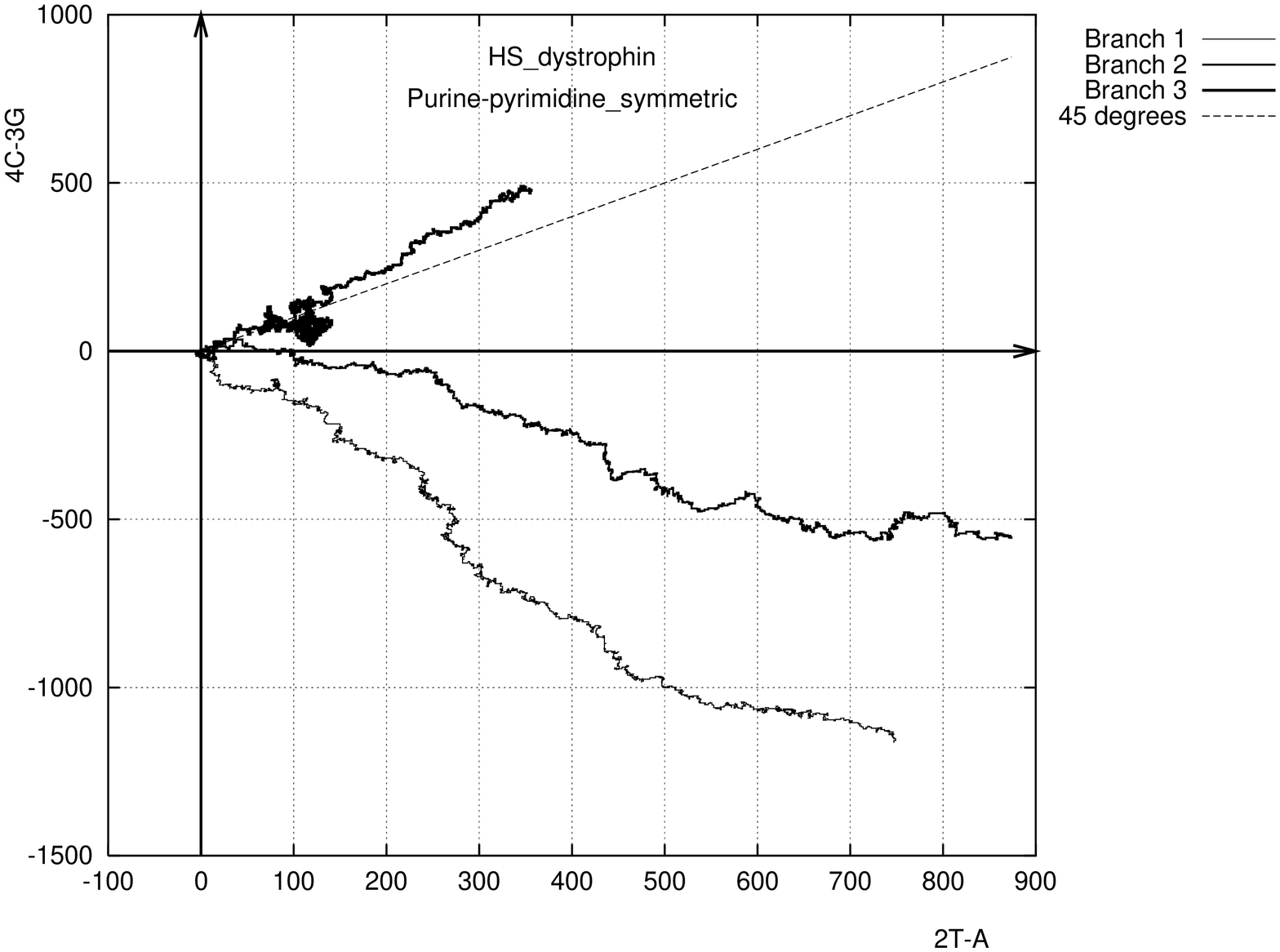}}
\centerline{\includegraphics[height=8cm,angle=-90]{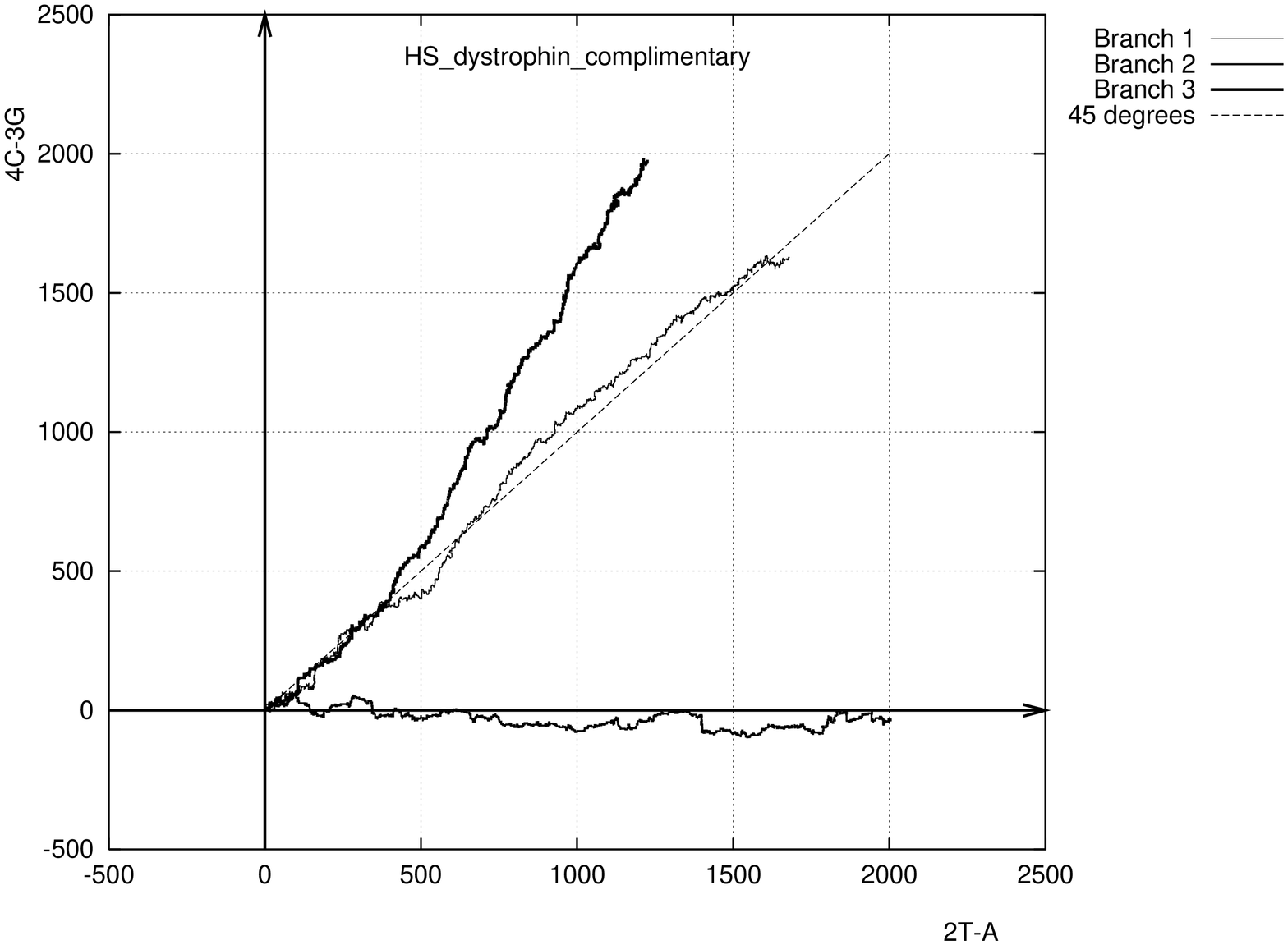}
\includegraphics[height=8cm,angle=-90]{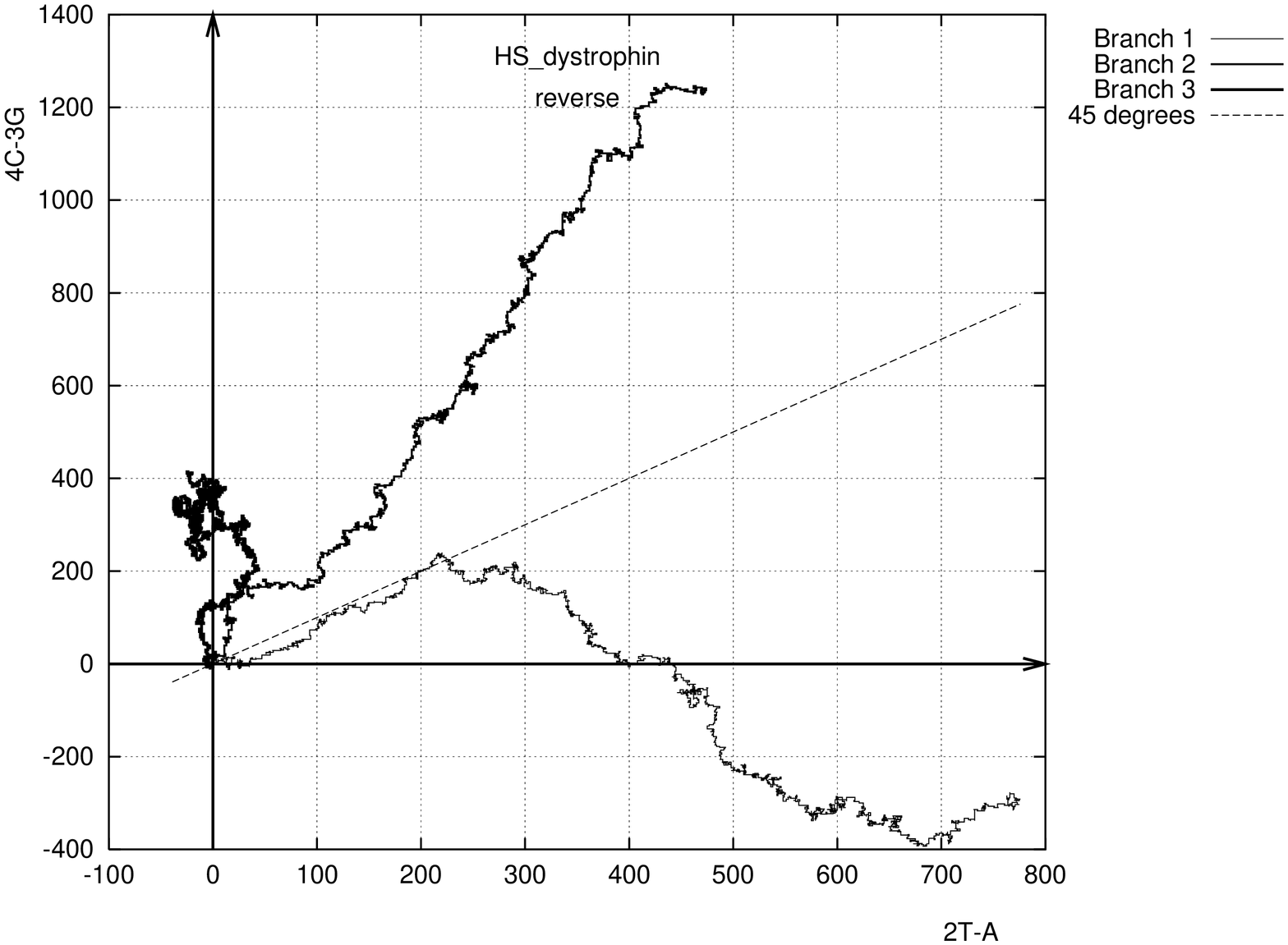}}\caption{The trianders
for the transformed sequences of Homo sapiens dystrophin (DMD), transcript
variant D140ab, mRNA in case of the amino-keto symmetry, purine-pyrimidine
symmetry, complementarity symmetry and reverse sequence reading. The original
nontransformed triander is shown on Fig. \ref{f-tri}.}%
\label{f-trans}%
\end{figure}

Here we illustrate how the symmetry transformations influence on the triander.
As an example we take the Homo sapiens dystrophin from Fig. \ref{f-tri}, and
result of various symmetry transformations (\ref{sym-p})--(\ref{sym-c}) and
reversing the sequence is shown on Fig. \ref{f-trans}.

We observe that the reverse triander is very similar to the original one on
Fig. \ref{f-tri}.

\subsection{Three-dimensional trianders}

The previously constructed two-dimensional trianders have the disadvantage
form, because it is not clear, where in the sequence a given point is. To
improve this we introduce three-dimensional trianders which are defined by the formula%

\begin{align}
x_{i}^{3D}  &  =\mathbf{d}_{\mathbf{T}}n_{\mathbf{T}}\left(  i\right)
-\mathbf{d}_{\mathbf{A}}n_{\mathbf{A}}\left(  i\right)  ,\label{3d-1}\\
y_{i}^{3D}  &  =\mathbf{d}_{\mathbf{C}}n_{\mathbf{C}}\left(  i\right)
-\mathbf{d}_{\mathbf{G}}n_{\mathbf{G}}\left(  i\right)  ,\label{3d-2}\\
z_{i}^{3D}  &  =i, \label{3d-3}%
\end{align}
which can be treated as mixing of one-dimensional and two-dimensional cases
with taking into account the determinative degree. Then, any on the DD space
structure can be definitely visually localized using vertical
axis.\begin{figure}[h]
\centerline{\includegraphics[height=8cm]{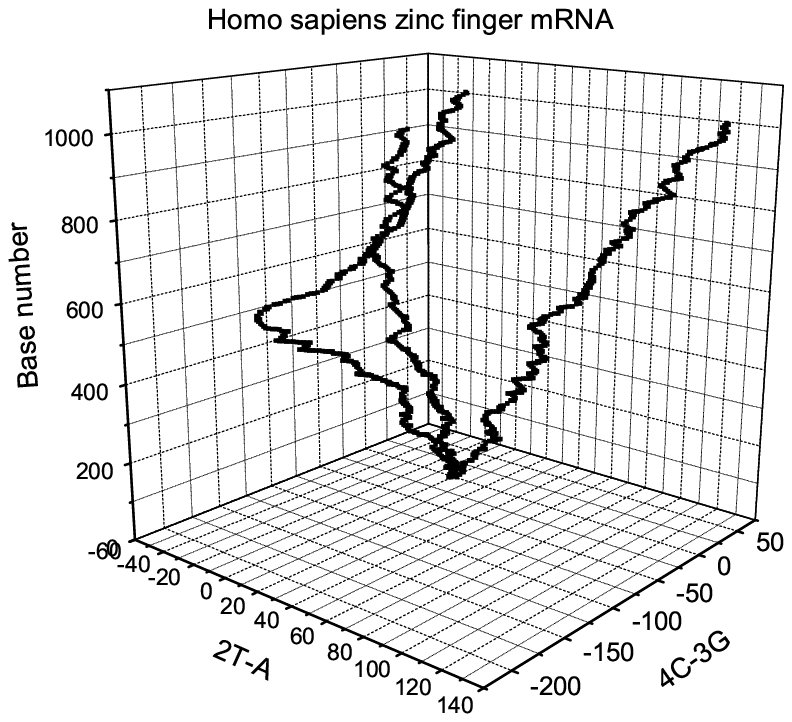}
\includegraphics[height=8cm]{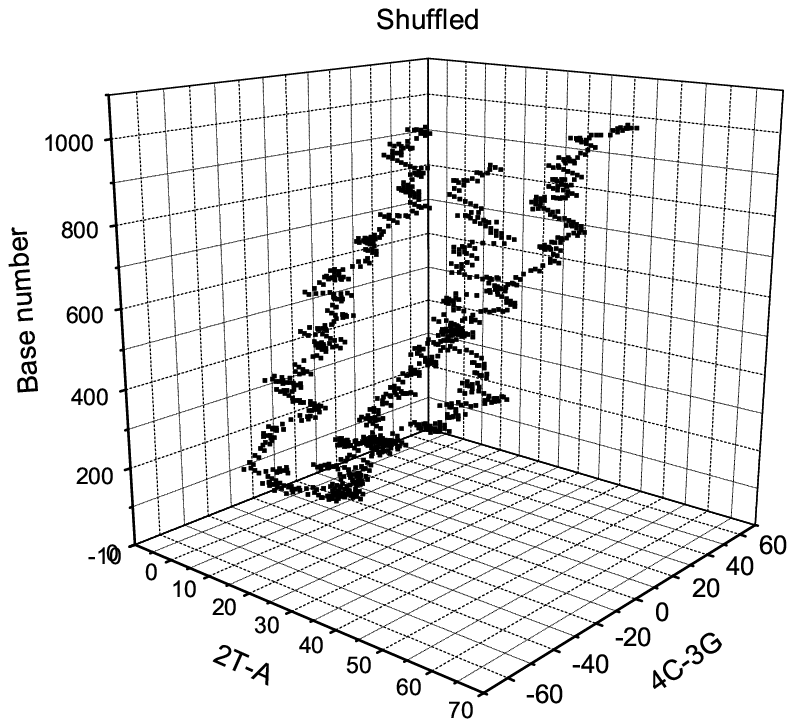}}\caption{Three-dimensional
trianders of the Homo sapiens zinc finger protein 265 (ZNF265), mRNA (left)
and its shuffled version (right).}%
\label{f-3d}%
\end{figure}
On the Fig. \ref{f-3d} we show
the three-dimensional triander of the Homo
sapiens zinc finger and its shuffled version.
All the graphs start from one point, the origin, and have different
length (which can be simply calculated from (\ref{3d-1})--(\ref{3d-3})),
characterizing them as a whole.

\section{Topological classification of trianders}

Here we propose the topological classification of trianders by their branches
belonging to different quadrants on the DD plane and to number of
intersections and return point. This makes possible studying the fine
structure of any length sequences with exactly established functions (genes,
intergenic space, repeat regions etc...) and comparing various locuses, as
well as searching for homological regions, which can allow us to work out
mathematically strong genomic signature formalism \cite{wu1,ber/ihm/bar}.

We note that there exist many types of trianders. A triander corresponding to
a gene we call a \textit{genogram}, and a triander corresponding to intergenic
space we call a \textit{gapgram.} If some branches intersect each other we say
about \textit{intersecting triander}, if a branch intersects itself producing
knots, we say about \textit{knot triander}. Branches can also have (multiple)
return points, and then we say about \textit{returned triander}. Thus, the
determinative degree\ walk \textquotedblleft topologizing\textquotedblright%
\ in our sense means that we identify trianders having definite structure
topological features (knot, intersection, return point) and place them into a
special class.

So we may hope that such topological classification of trianders can actually
help in solving by visual way the inverse problem: for a given sequence to
predict its possible function.

Let $n_{\mathbf{X}}\left(  i\right)  $ be cumulative quantity of nucleotide
$\mathbf{X}$ after $i$ steps, then the DD plane quadrants are defined by
(\ref{x})--(\ref{y}), and therefore

\textbf{I}: $\mathbf{2}n_{\mathbf{T}}\left(  i\right)  -n_{\mathbf{A}}\left(
i\right)  >0;\mathbf{4}n_{\mathbf{C}}\left(  i\right)  -\mathbf{3}%
n_{\mathbf{G}}\left(  i\right)  >0;$

\textbf{II}: $\mathbf{2}n_{\mathbf{T}}\left(  i\right)  -n_{\mathbf{A}}\left(
i\right)  <0;\mathbf{4}n_{\mathbf{C}}\left(  i\right)  -\mathbf{3}%
n_{\mathbf{G}}\left(  i\right)  >0;$

\textbf{III}: $\mathbf{2}n_{\mathbf{T}}\left(  i\right)  -n_{\mathbf{A}%
}\left(  i\right)  <0;\mathbf{4}n_{\mathbf{C}}\left(  i\right)  -\mathbf{3}%
n_{\mathbf{G}}\left(  i\right)  <0;$

\textbf{IV}: $\mathbf{2}n_{\mathbf{T}}\left(  i\right)  -n_{\mathbf{A}}\left(
i\right)  <0;\mathbf{4}n_{\mathbf{C}}\left(  i\right)  -\mathbf{3}%
n_{\mathbf{G}}\left(  i\right)  <0.$

After examination of around 2000 eukaryotic and prokaryotic sequences we found
all trianders can be distinguished into several types. The first type is a
\textit{chaotic triander} which has no definite branch structure, other ones
can be called \textit{ordered trianders}. To work out the general
classification of ordered trianders and description of branches we introduce
the notion:
\begin{equation}
\text{Type \textbf{A-B-C}}_{\mathbf{F}}^{\left(  x,y\right)  }%
\text{\ (\textbf{E})}, \label{t}%
\end{equation}
where \textbf{A} is quadrant where 1st branch lies, \textbf{B} is quadrant
where 2nd branch lies, \textbf{C} is quadrant where 3rd branch lies;
\textbf{E} is characteristics of triander as a whole; indices $\mathbf{F}$ and
$\left(  x,y\right)  $ describe properties of corresponding separate branches
(also for $\mathbf{A}$ and $\mathbf{B}$) which will be explained below.

In general there are $4^{3}=64$ possible ordered triander types classified by
quadrants only. We will identify trianders which differ by permutation,
because it corresponds to ORF shift, thus decreasing to 24 types.
Nevertheless, observation showed that there exist only 7 triander types:
\textbf{I-I-I}, \textbf{I-I-II}, \textbf{I-I-III}, \textbf{I-I-IV},
\textbf{I-II-III}, \textbf{I-II-IV}, \textbf{I-III-IV}. For example, the Type
\textbf{I-I-II} includes the Types \textbf{I-II-I} and Types \textbf{II-I-I},
if we shift ORF to 1 and 2, but on figures we show exact triander names
(\ref{t}).

If e.g. a branch crosses from \textbf{I }quadrant to \textbf{II }quadrant, we
denote that by fraction \textbf{I/II}. For instance, the triander of Homo
Sapiens dystrophin gene Fig. \ref{f-tri}, is of Type \textbf{II-I-I/IV}.

The additional qualitative features of triander as a whole observed from
sequence examination are
\[
\text{\textbf{E}=sharp, flat, parallel.}%
\]
For branch properties we have
\begin{gather*}
\text{x,y}\ \text{denotes axis to which a branch is parallel,}\\
\text{\textbf{F}=blury, loop, smooth, oscillative (horizontal, vertical).}%
\end{gather*}

Separately we can describe \textquotedblleft interaction\textquotedblright\ of
branches as:

1) Single intersection of $\mathbf{A}$ and $\mathbf{B}$ is denoted by sign
$\mathbf{A\#B}$, which gives \textit{intersecting triander};

\begin{figure}[tbh]
\centerline{\includegraphics[height=8cm,angle=-90]{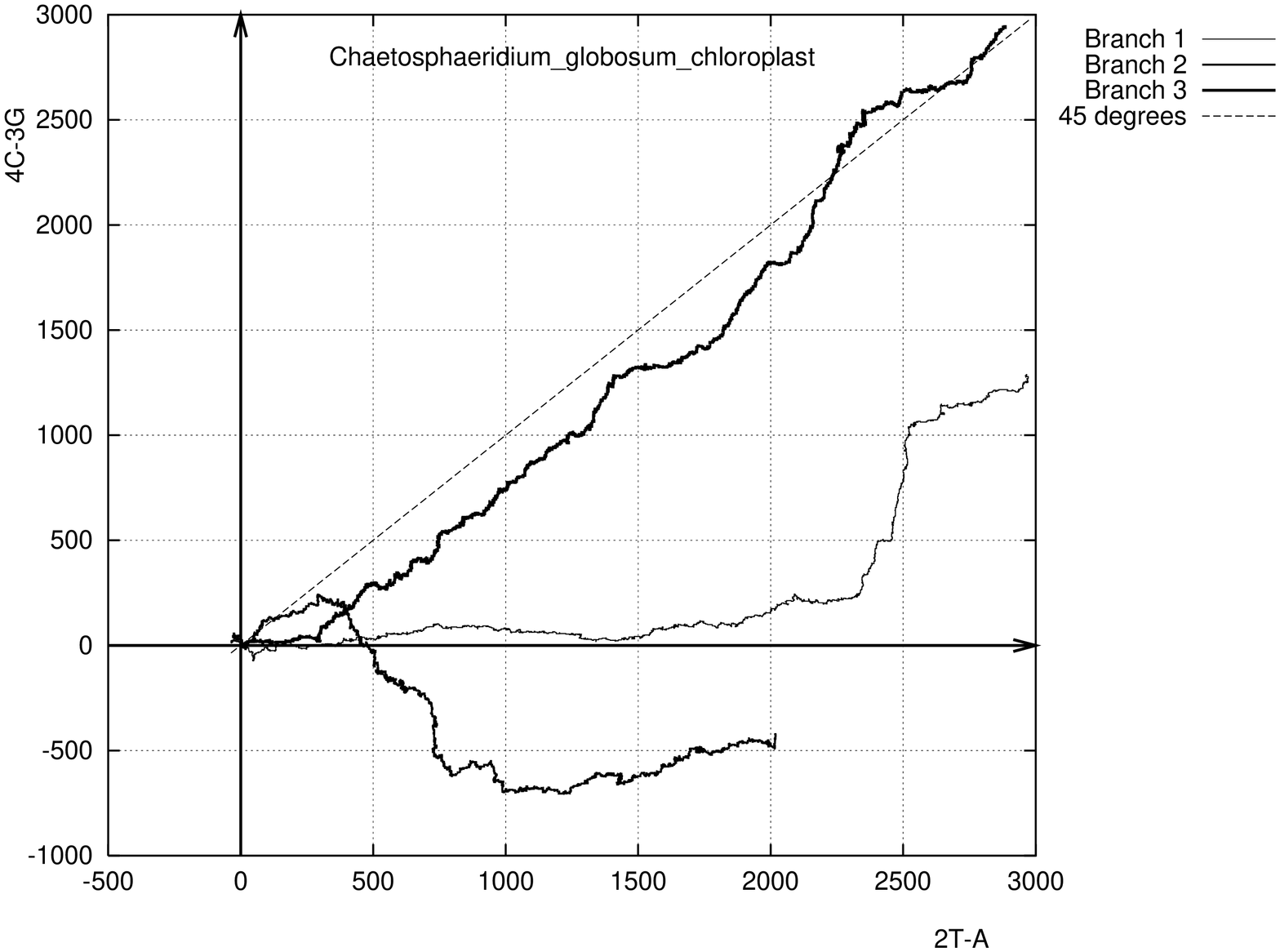}
\includegraphics[height=8cm,angle=-90]{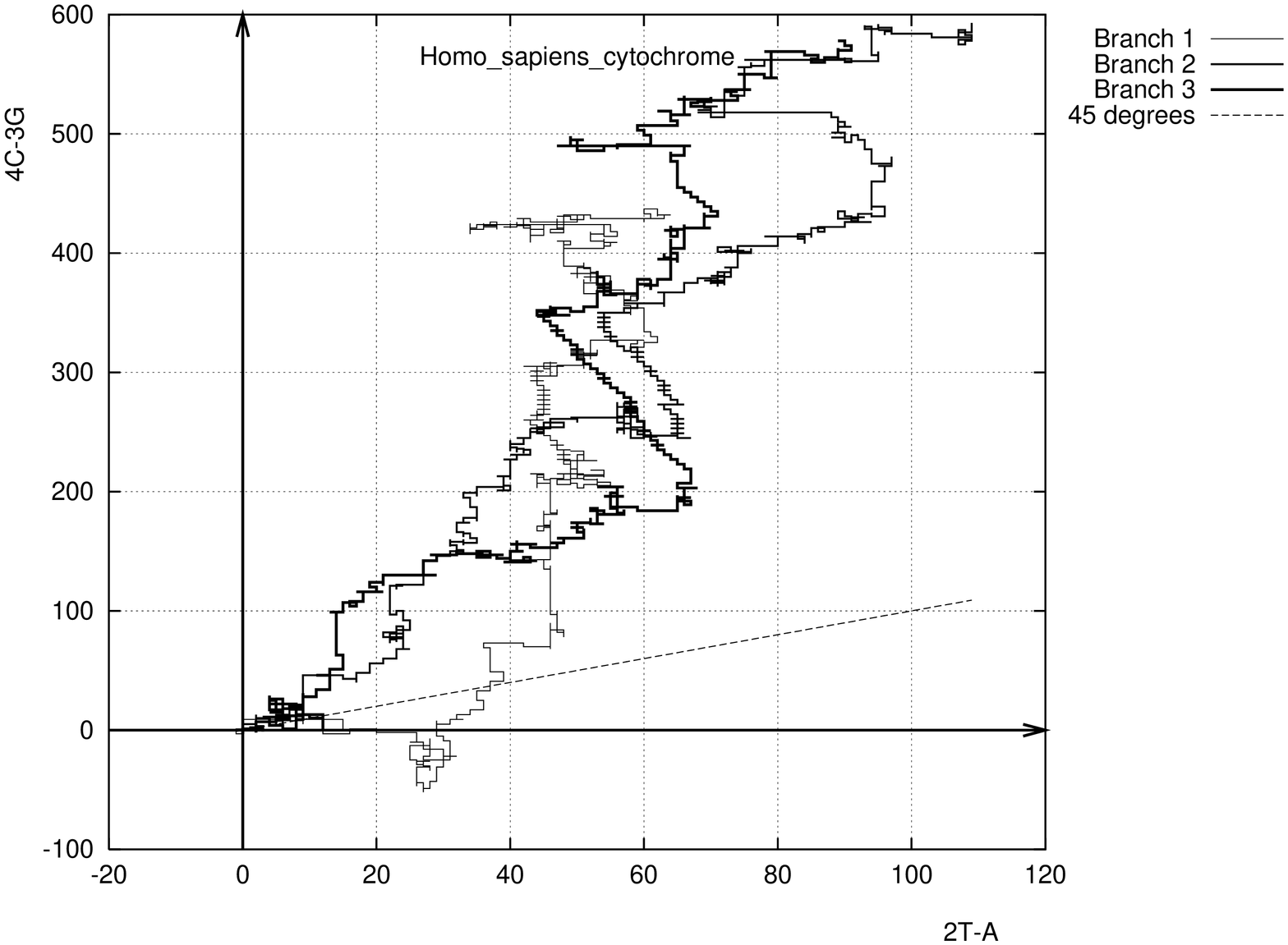}}\caption{Intersecting
triander, Chaetosphaeridium globosum chloroplast, complete genome (left) and
braiding triander Homo sapiens cytochrome P450 2f1 (CYP2F1P) (right).}%
\label{f-bri}%
\end{figure}

2) Multiple intersection $\mathbf{A}$ and $\mathbf{B}$ is called braiding and
denoted $\mathbf{A}\between\mathbf{B}$, which gives \textit{braiding
triander}. See Fig. \ref{f-bri}.

We have thoroughly analyzed 150 sequences different by function and evolution
level, and for each sequence there were also constructed 100 shuffled
sequences having the same nucleotide composition, but not coinciding with the
examined one. They are presented in the Tables \ref{tabl1},\ref{tabl2}.

\begin{table}[ptb]
\caption{Examined sequences}%
\label{tabl1}
\par
\begin{center}%
\begin{tabular}
[c]{|l||l|c|}\hline
Type & Sequence description & Number\\\hline\hline
& Arabidopsis thaliana clone 7867 mRNA, & \\
& C.elegans essential lethal-805, myotactin complex & 2\\
& C.elegans heterochronic gene LIN-42 & 1\\
& C.elegans ribosomal protein L32 & 1\\
I,I,I & Dros melanogaster chromosome 3R & 1\\
& HS apoptosis-associated tyrosine kinase & 2\\
& HS coagulation factor VIII & 1\\
& HS COL4A6 gene for a6(IV) collagen & 2\\
& HS cytochrome P450,family 1-8, subfamily A,B,C,F,J & 15\\
& HS DNA cross-link repair 1A,1C (PSO2 homolog, S. cerevisiae) & 2\\
& HS dystrophin (DMD) gene, exons,transcript variants & 4\\
& HS exportin, tRNA & 1\\
& HS FGG gene for fibrinogen, & 2\\
& HS fibrinogen alpha chain gene, complete mRNA & 4\\
& HS glutathione synthetase (GSS), mRNA & 2\\
& HS H19 gene,complete sequence,mRNA & 2\\
& HS myeloid ecotropic viral integration site mRNA & 1\\
& HS MESTIT1 antisense RNA,intergenic space & 1\\
& HS mRNA 10q24.3-qter & 1\\
& Hs mucosal vascular addressin cell adhesion molecule & 1\\
& HS phenylalanine hydroxylase (PAH) & 3\\
& Hs phytanoyl-CoA hydroxylase interacting protein & 1\\
& HS retinal dystrophin (DMD) gene Exon 30 & 1\\
& Hs solute carrier family 22 (organic anion transporter), & 1\\
& HS suppressor of cytokine signaling,intergenic space & 1\\
& HS syntrophin, alpha 1 & 3\\
& HS vitelliform macular dystrophy & 2\\
& HS, chorionic somatotropin hormone & 1\\
& HS,genomic DNA chrom1,intregenic space & 1\\
& HS,genomic DNA P450 intergenic space & 1\\
& Human CYP2D7BP pseudogene for cytochrome P450 2D6 & 1\\
& Human mRNA encoding placental lactogen hormone & 1\\
& Human nested gene protein gene & 1\\
& Mus musculus insulin-like growth factor & 2\\
& Mus musculus interleukin 11 receptor & 1\\
& Mus musculus like-glycosyltransferase mRNA & 1\\
& Mus musculus similar to mitochondrial ribosomal protein S36 & 2\\
& Rat gene for alpha-fibrinogen & 1\\
& Rattus norvegicus cytochrome P450 IIA3 mRNA, 3' end & 2\\
& Similar to FBJ murine osteosarcoma viral oncogene & 1\\
& Takifugu rubripes DMD gene & 1\\\hline\hline
\end{tabular}
\end{center}
\end{table}

\begin{table}[ptb]
\caption{Examined sequences}%
\label{tabl2}
\begin{center}%
\begin{tabular}
[c]{|l||l|c|}\hline
Type & Sequence description & Number\\\hline\hline
& Dengue virus type 1 strain FGA/NA,complete genome. & 1\\
& HS ATP-binding cassette, sub-family C (CFTR/MRP),mRNA & 3\\
& HS aldehyde dehydrogenase 1 family & 5\\
& Human fibrinogen beta-chain mRNA, partial cds 41b1,short asym & 2\\
& Homo sapiens fibrinogen-like 1 (FGL1), transcript variant & 1\\
& C.elegans immunoglobulin & 1\\
& Hs spondyloepiphyseal dysplasia late RNA & 1\\
& C.elegans nematode cuticle collagen & 1\\
& C.elegans cuticle collagen family member (28.9 kD) & 1\\
I,I,IV & Chaetosphaeridium globosum chloroplast, complete genomes & 1\\
& C. elegans Collagen with Endostatin domain CLE-1 & 1\\
& Hs aldehyde dehydrogenase 1 family, member A2 (ALDH1A2) & 1\\
& Hs chromosome 20 open reading frame 1 (C20orf1), mRNA & 1\\
& HS ATP-binding cassette, sub-family C (CFTR/MRP)mRNA & 2\\
& Human cytochrome P450 (CYP2A13) gene, complete cds. & 1\\
& Hs cytochrome P450 2S1 (CYP2S1) mRNA, complete cds. & 1\\
& Hs P450 (cytochrome) oxidoreductase (POR), & 1\\
& Hs cytochrome P450, family 2,39,51,20 & 4\\
& Hs chromosome 1 MRG1 intergenic space & 1\\
& Hs cytochrome P450 intergenic space & 1\\
I,II,IV & HS dystrophin (DMD) D140ab, variants,mRNA. & 5\\
& H.sapiens mRNA for ribosomal protein L30 & 1\\
& Human mRNA for ribosomal protein L32 & 1\\\hline
I,I,III & Hs collagen, type IX, alpha 2 (COL9A2), mRNA & 1\\
& Mus musculus ribosomal protein L32 & 1\\
& Mus musculus, ribosomal protein L30 BC002060 & 1\\
& Homo sapiens apoptosis antagonizing transcription factor (AATF) & 5\\\hline
I,II,III & Homo sapiens H1F5 histone family & 1\\
& Hs cytochrome P450, family 17, subfamily A, polypeptide 1 & 1\\
& HS genomic cluster,H1 histone family, member 5 & 1\\
& HS survival of motor neuron 1, telomeric (SMN1) & 2\\
& HS Che-1 mRNA, complete cds.123mRNA & 1\\
& Hs ZIS1protein 265 (ZNF265) mRNA, & 1\\\hline
I,III,IV & Hs utrophin (homologous to dystrophin)3bls11 & 1\\
& Macaca fascicularis RPL30 mRNA, family & 4\\\hline
& HS bestrophin (VMD2) mRNA, alternatively spliced product & 1\\
I,I,II & HS BTG family, member 2 (BTG2), mRNA & 1\\
& HS Cbp/p300-interacting transactivator,with Glu/Asp-rich & 1\\
& Human msg1-related gene 1 (mrg1) mRNA & 1\\\hline
\end{tabular}
\end{center}
\end{table}

\begin{figure}[tbh]
\centerline{\includegraphics[height=8cm,angle=-90]{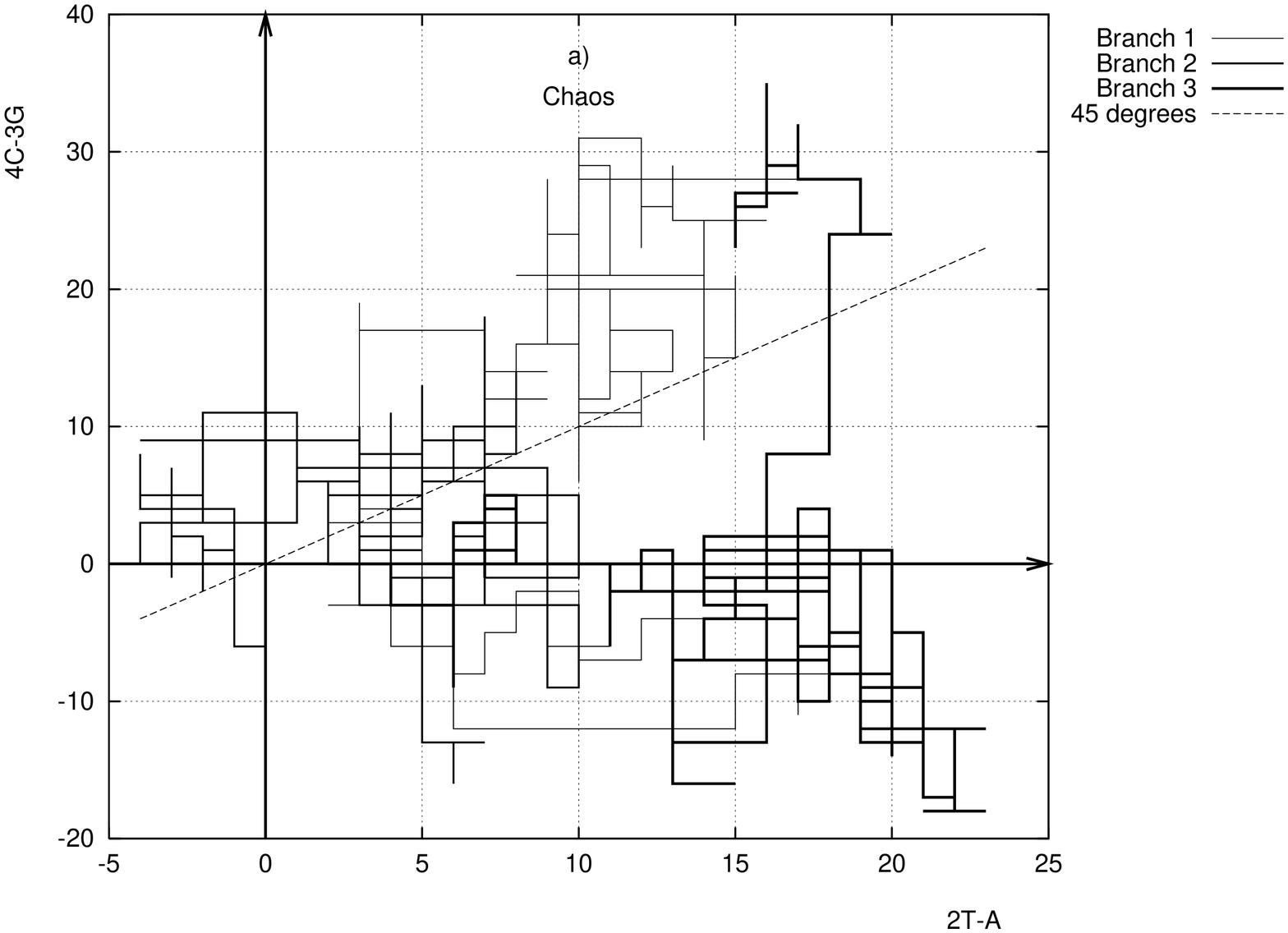}
\includegraphics[height=8cm,angle=-90]{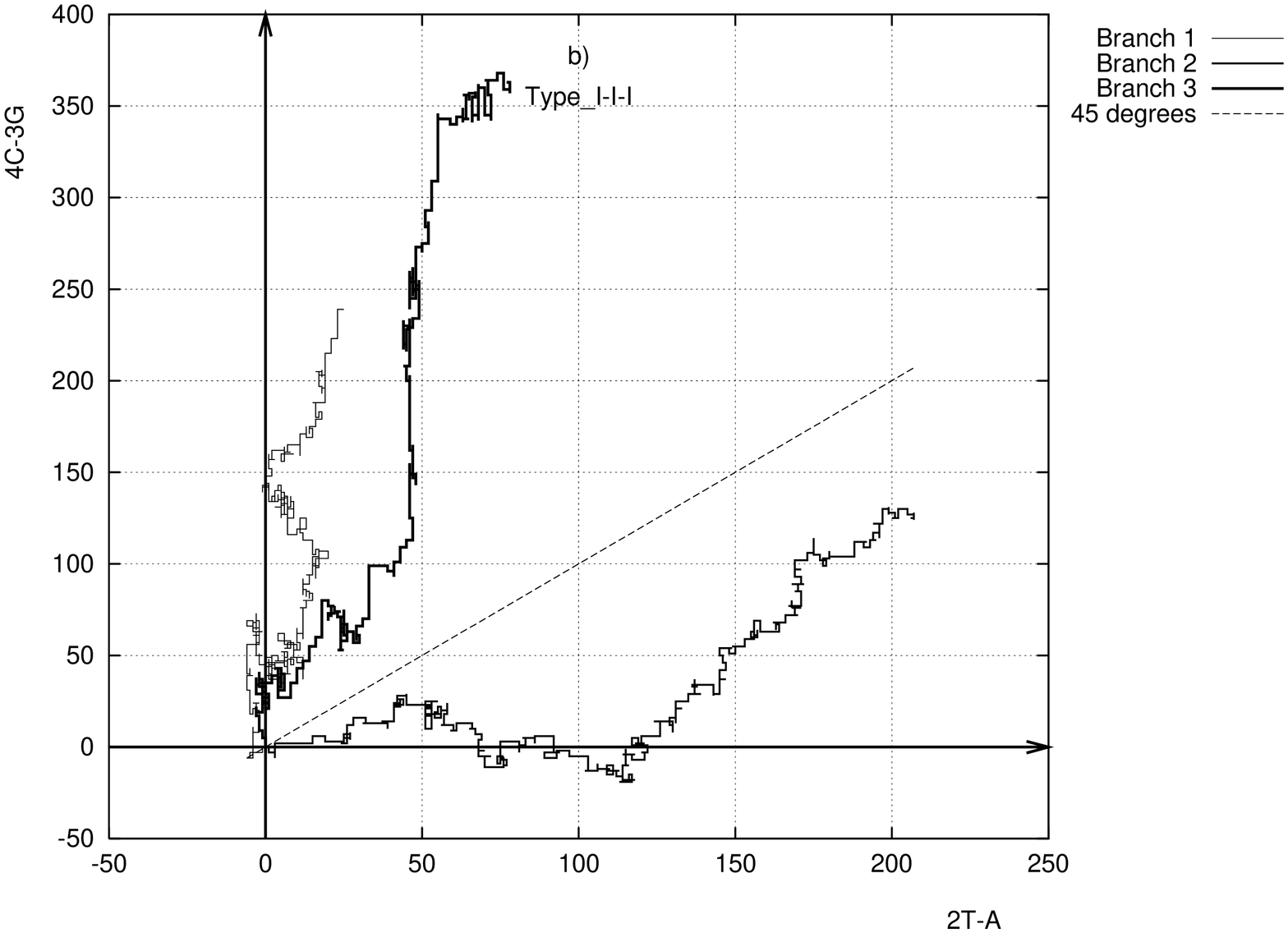}}
\centerline{\includegraphics[height=8cm,angle=-90]{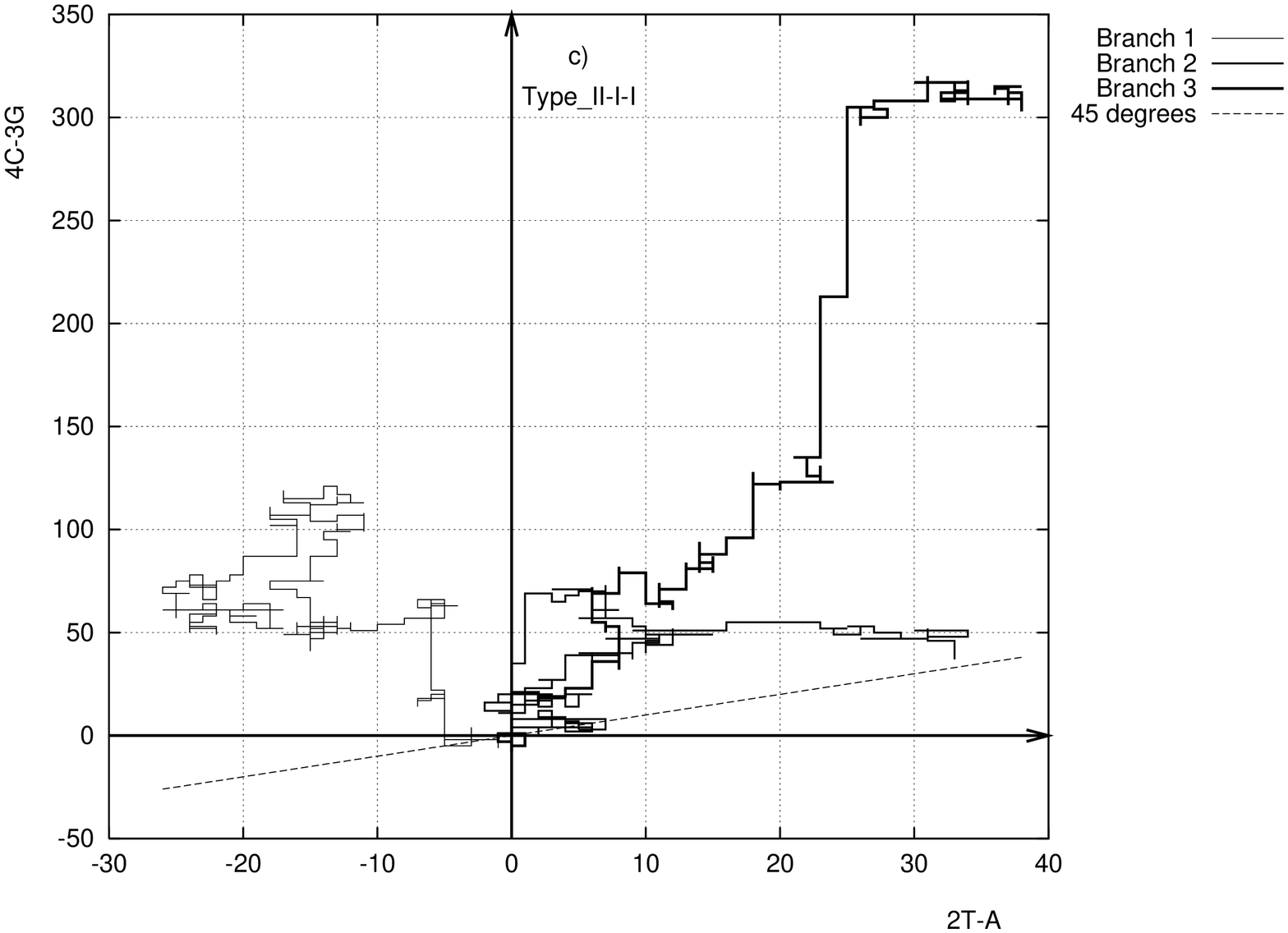}
\includegraphics[height=8cm,angle=-90]{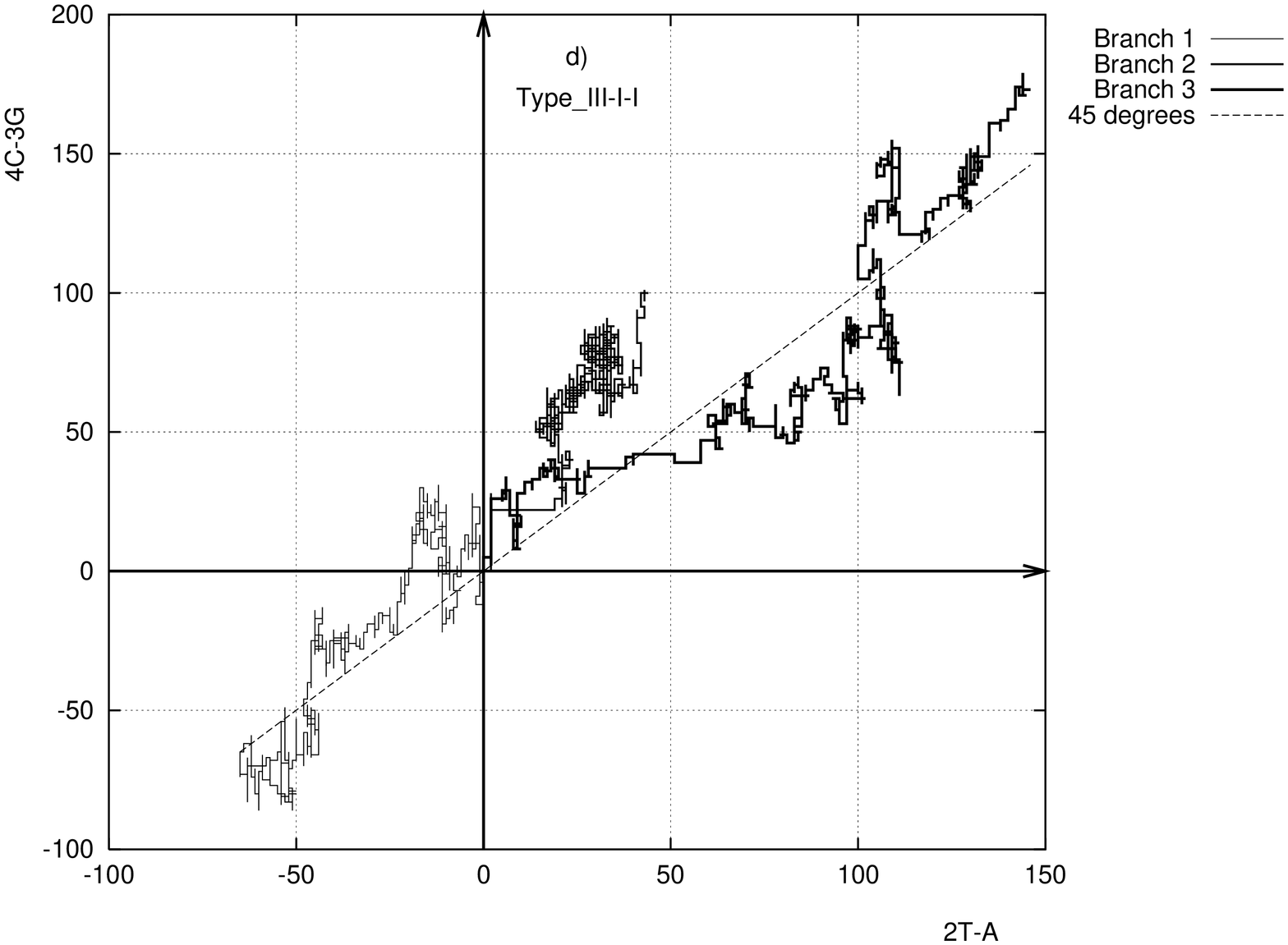}}
\centerline{\includegraphics[height=8cm,angle=-90]{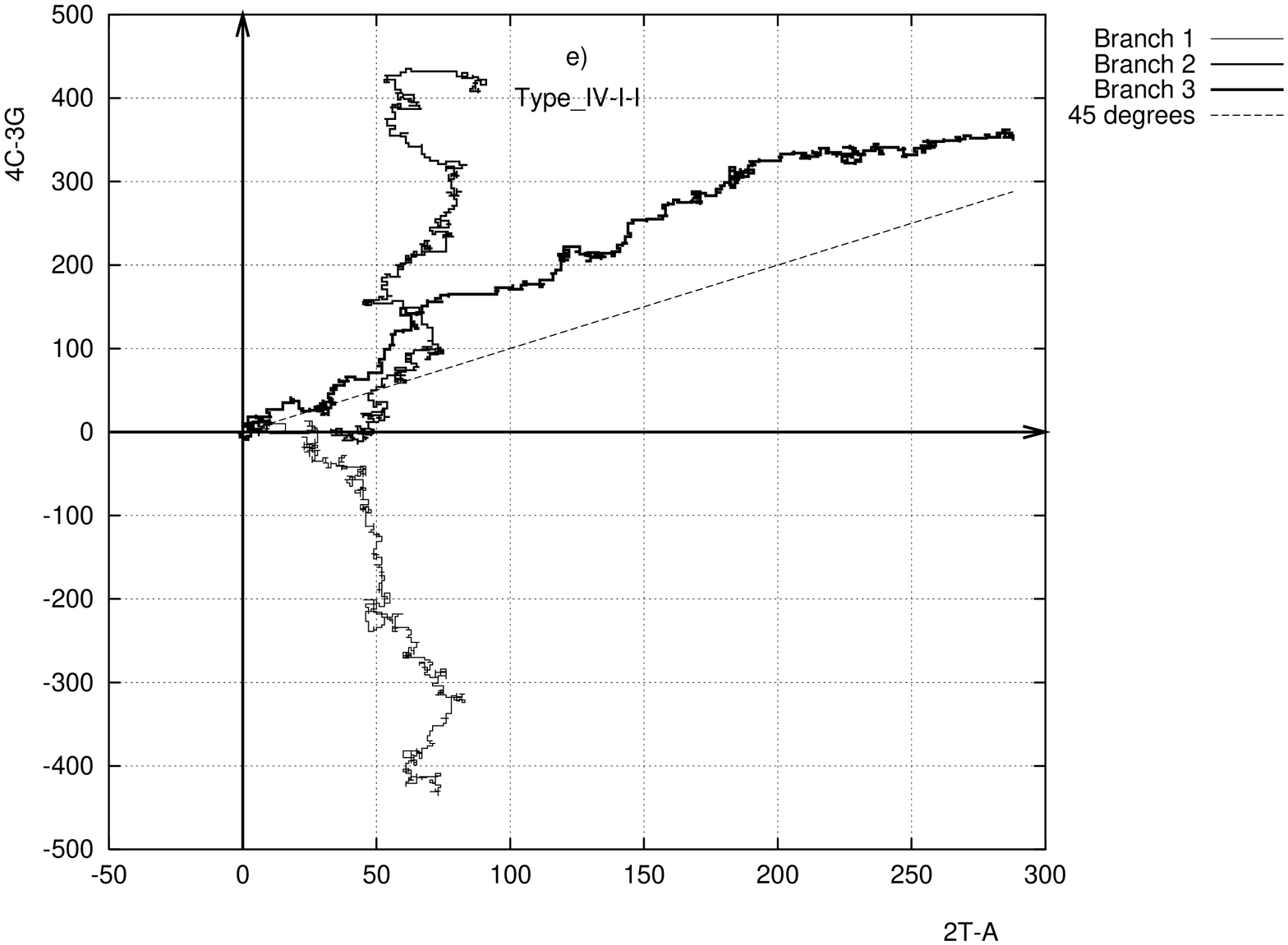}
\includegraphics[height=8cm,angle=-90]{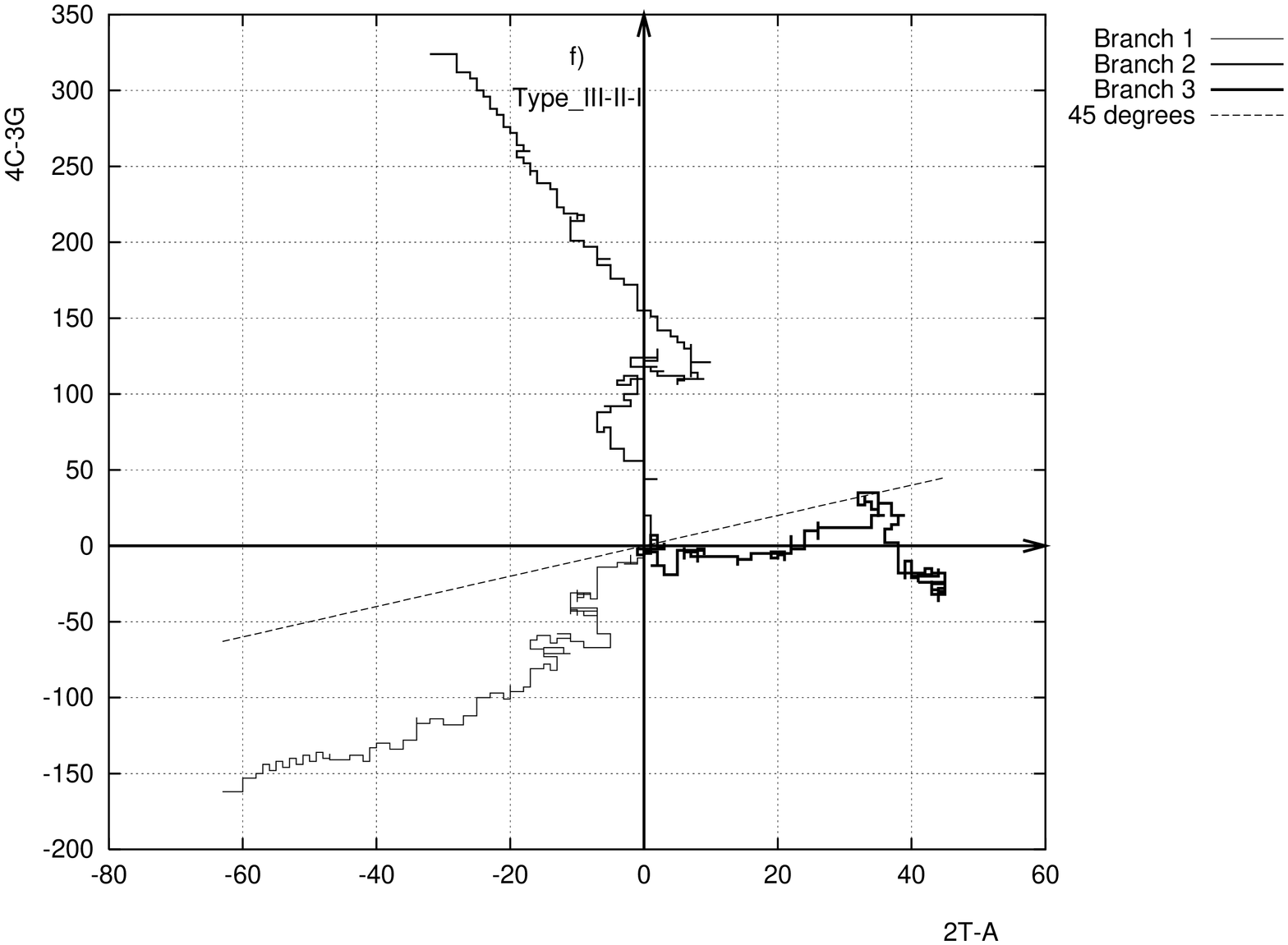}}
\centerline{\includegraphics[height=8cm,angle=-90]{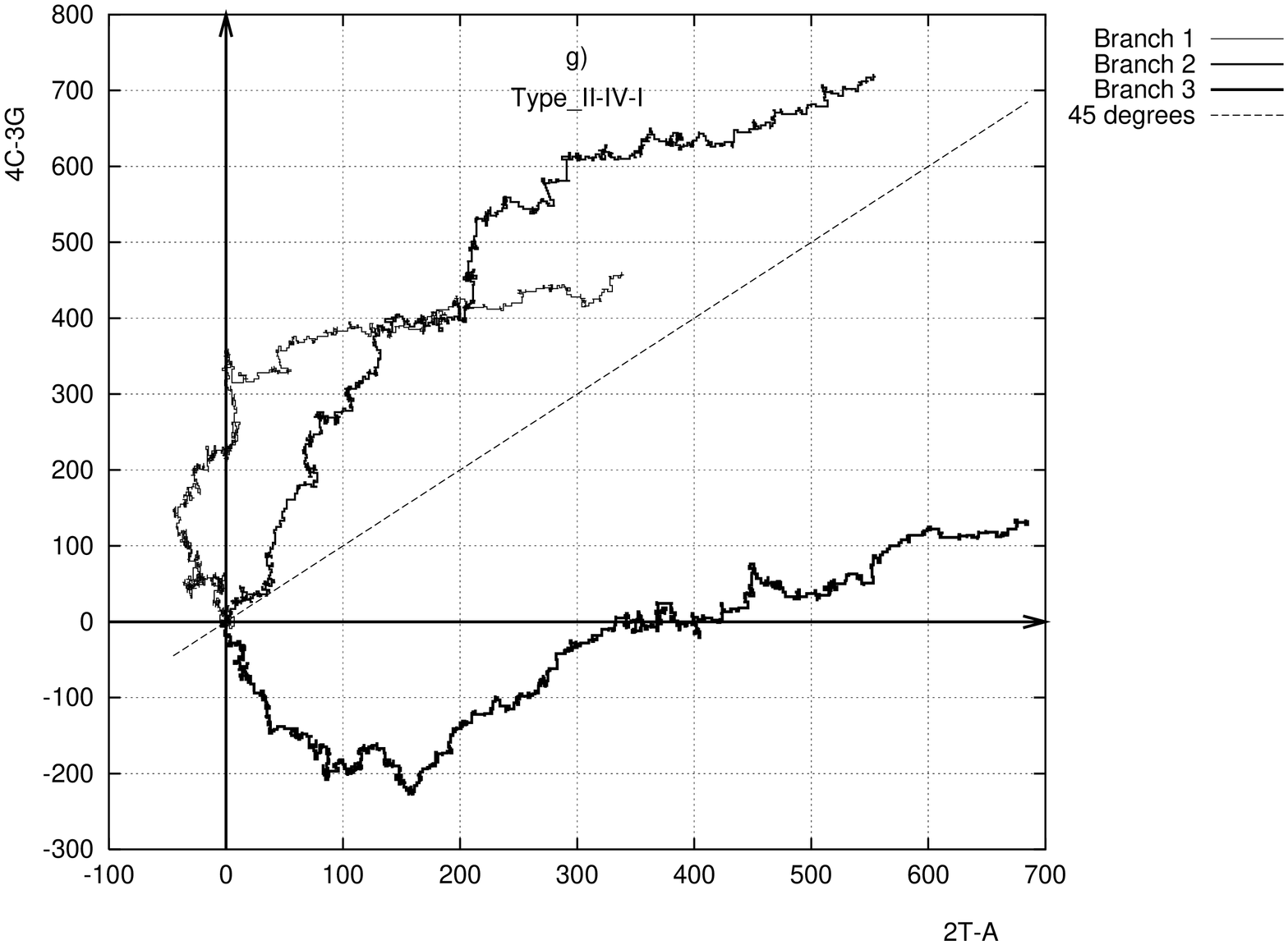}
\includegraphics[height=8cm,angle=-90]{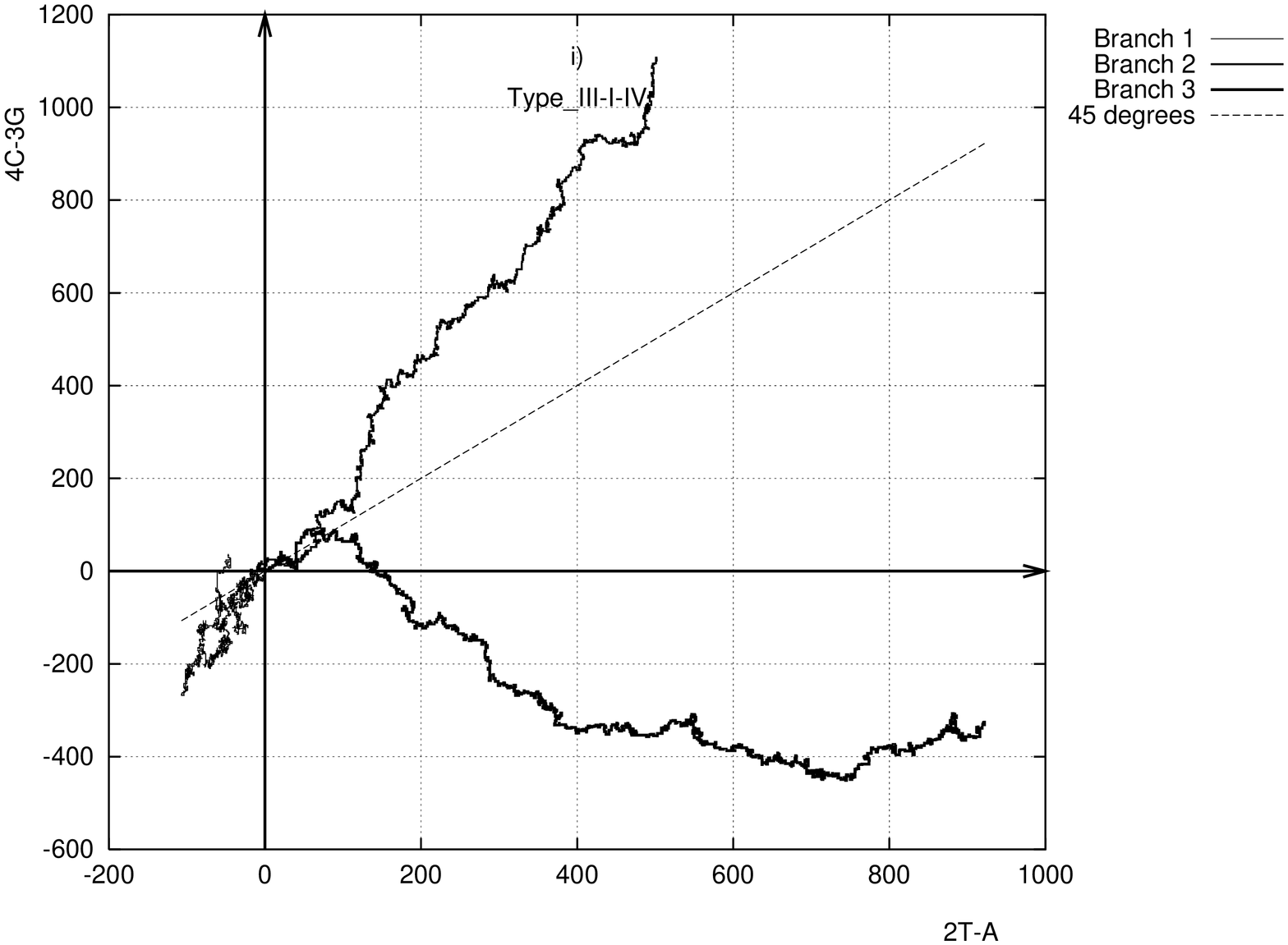}}\caption{Classification of
trianders.}%
\label{f-all}%
\end{figure}

For every class we show a typical triander of Fig. \ref{f-all}, where the
following real sequences are presented:

\textbf{a)} Chaotic triander. Dengue virus type 1 strain FGA/NA d1d,
intergenic space: AF226686.

\textbf{b)} Type \textbf{I-I-I}$^{y}$. Homo sapiens cytochrome P450, family 2,
subfamily F, polypeptide 1 (CYP2F1), mRNA: NM\_000774.

\textbf{c)} Type \textbf{II}$_{blury}$\textbf{-I}$^{x}$\textbf{-I}$^{y}%
$(flat). Homo sapiens Cbp/p300-interacting transactivator, mRNA: NM\_006079.

\textbf{d)} Type \textbf{III}$_{oscill}$\textbf{-I}$_{blury}$\textbf{-I}%
$_{oscill}$. Homo sapiens collagen, type IX, alpha 2 (COL9A2), mRNA: NM\_001852.

\textbf{e)} Type \textbf{IV-I\#I}. Caenorhabditis elegans immunoglobulin
domain-containing protein family member (106.4 kD), mRNA: NM\_171617.

\textbf{f)} Type \textbf{III-II-I}$^{x}$(sharp). Homo sapiens H1 histone
family, member 5 (H1F5), mRNA: NM\_005322.

\textbf{g)} Type \textbf{II\#I-IV}. Homo sapiens dystrophin (muscular
dystrophy, Duchenne and Becker types) (DMD), transcript variant D140ab, mRNA: NM\_004022.

\textbf{h)} Type \textbf{I}$_{loop}$\textbf{-I-IV}. Homo sapiens utrophin
(homologous to dystrophin) (UTRN), mRNA: NM\_007124.

Further more careful topological classification and analysis of two- and
three-dimensional trianders can be made using some of the topological curves
methods \cite{pet,rok,arn/ole} or the knot theory (see e.g.
\cite{turaev,kauffman}).

\section{Conclusions}

We can conclude that the introduced determinative degree DNA\ walk method
confirms the \textquotedblleft mosaic\textquotedblright\ stricture of genome,
shows parts with different nucleotide content and \textquotedblleft
strength\textquotedblright, and so allows us to find the \textquotedblleft
fine structure\textquotedblright\ of nucleotide sequences.

We propose a general method of identification of DNA\ sequence
\textquotedblleft by triander\textquotedblright, which can be treated as a
unique \textquotedblleft genogram\textquotedblright, \textquotedblleft gene
passport\textquotedblright, etc. The two- and three-dimensional trianders are
introduced and their features are studied.

The difference of the nucleotide sequences fine structure in genes and the
intergenic space is shown. Also there is a clear triplet signal in coding
locuses which is absent in the intergenic space and is independent from the
sequence length, but depends from composition only. All plots are compared
with corresponding shuffled sequences of the same nucleotide composition,
which allows us to extract real ordering effect from composition influence.

We have constructed the classification of trianders, on its basis a detail
working out signatures of functionally different genomic regions can be made.

\medskip

\textbf{Acknowledgments}. We would like to thank S. Cebrat and M. R. Dudek for
kind hospitality at SmORFland (Inst. Microbiology, Wroclaw) and sharing their
useful DNA walk experience. Also we are grateful to A. Yu. Berezhnoy, N. A.
Chashchin, A. A. Gusakov, V. Knecht, G. Ch. Kourinnoy, P. Mackiewicz, B. V.
Novikov, L. A. Livshits, S. Sachse, Yu. G. Shckorbatov, O. A. Tretyakov for
very fruitful discussions, and G. Findley, H. Grubm\"{u}ller, P. Jarvis, D. C.
Torney, C. Zhang for sending their interesting papers. We are thankful to S.
M. Donets for language checking and V. Kalashnikov for creation of original
special programs for sequence analysis and indispensable assistance in the
related GNUPLOT/PERL/C programming.

\end{document}